%% file: main.tex
\documentclass[conference]{IEEEtran}

\input{Include/packages}

\input{Include/cmds}

\input{Include/shorthand}
\usepackage{placeins}
\def\FIGDIR{./Figures}
\def\BibTeX{{\rm B\kern-.05em{\sc i\kern-.025em b}\kern-.08em
    T\kern-.1667em\lower.7ex\hbox{E}\kern-.125emX}}

\begin{document}

%% EDIT TITLE BELOW

\title{Understanding the Performance Horizon of
the Latest ML Workloads with NonGEMM Workloads}
% \title{Understanding the Performance Horizon of the Latest ML Workload with Non-GEMM Workloads: An Empirical Approach}

%% DO NOT EDIT THE FOLLOWING

%\renewcommand\Authsep{\qquad}
%\renewcommand\Authand{\qquad}
%\renewcommand\Authands{\qquad}

%% EDIT AUTHOR LIST BELOW

\author{
\IEEEauthorblockN{Rachid Karami}
\IEEEauthorblockA{\textit{Electrical Engineering and Computer Science} \\
\textit{University of California, Irvine}\\
Irvine, USA \\
karamir@uci.edu}
\and 
\IEEEauthorblockN{Sheng-Chun Kao}
\IEEEauthorblockA{ 
Mountain View, USA \\
chuchu40507@gmail.com}
\and 
\IEEEauthorblockN{ Hyoukjun Kwon}
\IEEEauthorblockA{\textit{Electrical Engineering and Computer Science} \\
\textit{University of California, Irvine}\\
Irvine, USA \\
hyoukjun.kwon@uci.edu}
}
% \author{Author2 Name}
% \author{Author2 Name}
% \author{Author2 Name}

% \author{Author3 Name}
% \affiliation{Full Name of Awesome School}

%%% ALTERNATIVE FORMAT FOR MULTIPLE SCHOOLS:
%%% 
% \author{
% }

\maketitle
% \thispagestyle{firstpage}
%\pagestyle{plain}

%% EDIT YOUR PAPER'S CONTENTS BELOW
\input{Sections/00_Abstract}
\input{Sections/01_Introduction}

\input{Sections/02_Background}
\input{Sections/03_Benchmark}
\input{Sections/04_Methodology}
\input{Sections/05_CaseStudies}
\input{Sections/06_RelatedWorks}

\input{Sections/07_Conclusion}

%%%%%%%%% -- BIB STYLE AND FILE -- %%%%%%%%
\bibliographystyle{IEEEtranS}
\bibliography{ref}
%%%%%%%%%%%%%%%%%%%%%%%%%%%%%%%%%%%%
\input{Sections/09_Appendix_A}

\end{document}

%% file: Include/packages.tex
\usepackage{cite}
\usepackage{amsmath,amssymb,amsfonts}
\usepackage{algorithmic}
\usepackage{graphicx}
\usepackage[dvipsnames]{xcolor}
\usepackage[final]{microtype}
\usepackage[italic]{mathastext}
\usepackage{libertine}
\usepackage[T1]{fontenc}
\usepackage{textcomp}
\usepackage[varqu,varl]{zi4}
\usepackage[all]{nowidow}
\usepackage[keeplastbox]{flushend}
\usepackage{fancyhdr}

\usepackage{color}
\usepackage{microtype}
\usepackage{xspace}
\usepackage{graphicx}
\usepackage{subfigure}
\usepackage{booktabs} % for professional tables

\usepackage{multirow}
\usepackage{comment}

\usepackage{hyperref}
\usepackage{bm}

\usepackage{todonotes}

\usepackage{pifont}

\usepackage{amsmath,amsfonts}
\usepackage{algorithmic}
\usepackage{graphicx}
\usepackage{textcomp}
\usepackage{xcolor}
\usepackage{multirow}
\RequirePackage[nameinlink]{cleveref} % 'nameinlink' to mimic \autoref's behavior
\crefname{chapter}{Chapter}{Chapters}
\crefname{section}{Section}{Sections}
\crefname{subsection}{Section}{Sections}
\crefname{equation}{Equation}{Equations}
\crefname{definition}{Definition}{Definitions}
\crefname{assumption}{Assumption}{Assumptions}
\crefname{theorem}{Theorem}{Theorems}
\crefname{figure}{Figure}{Figures}
\crefname{table}{Table}{Tables}
\crefname{algorithm}{Algorithm}{Algorithms}
\let\autoref\cref % set \autoref as an alias for \cref

%% file: Include/cmds.tex
\newcommand{\insertFigure}[2]{
    \begin{figure}[t]
        \centering
        \includegraphics[width=\linewidth]{\FIGDIR/#1.pdf}
        \vspace{-7mm}
        \caption{#2}
        \vspace{-4mm}
        \label{fig:#1}
    \end{figure}
}

\newcommand{\insertWideFigure}[2]{
    \begin{figure*}[h]
        \centering
        \includegraphics[width=\textwidth]{\FIGDIR/#1.pdf}
        \vspace{-6mm}
        \caption{#2}
        \label{fig:#1}
    \end{figure*}
}

\newcommand{\insertWideFigureShrink}[3]{
    \begin{figure*}[h]
        \centering
        \includegraphics[width=#3\textwidth]{\FIGDIR/#1.pdf}
        \caption{ #2}
        \label{fig:#1}
    \end{figure*}
}

\ifx\forsubmission\undefined
\newcommand{\TODO}[1]{\textcolor{red}{TODO: #1}}
\newcommand{\SK}[1]{\textcolor{red}{SK: #1}}
\newcommand{\HK}[1]{\textcolor{blue}{HK: #1}}
\newcommand{\RK}[1]{\textcolor{brown}{RK: #1}}

\newcommand{\fixme}[1]{{\color{red} {#1}}}

\else
\newcommand{\TODO}[1]{\textcolor{red}{}}
\newcommand{\SK}[1]{\textcolor{red}{}}
\newcommand{\FK}[1]{\textcolor{blue}{}}
\newcommand{\HK}[1]{\textcolor{blue}{}}
\newcommand{\RK}[1]{\textcolor{green}{}}

\newcommand{\fixme}[1]{{\color{red} {}}} 

\fi

\newcommand{\squishlist}{
 \begin{list}{$\bullet$}
  { \setlength{\itemsep}{0pt}
     \setlength{\parsep}{3pt}
     \setlength{\topsep}{3pt}
     \setlength{\partopsep}{0pt}
     \setlength{\leftmargin}{1.5em}
     \setlength{\labelwidth}{1em}
     \setlength{\labelsep}{0.5em} } }

\newcommand{\squishlisttwo}{
 \begin{list}{$\bullet$}
  { \setlength{\itemsep}{0pt}
     \setlength{\parsep}{0pt}
    \setlength{\topsep}{0pt}
    \setlength{\partopsep}{0pt}
    \setlength{\leftmargin}{2em}
    \setlength{\labelwidth}{1.5em}
    \setlength{\labelsep}{0.5em} } }

\newcommand{\squishend}{
  \end{list}  }

\newcommand{\betterparagraph}[1]{\noindent \textbf{#1. }}

%% file: Include/shorthand.tex
\newcommand{\bench}{\textsc{NonGEMM Bench}\xspace}

%% file: Sections/00_Abstract.tex
\begin{abstract}
%Machine Learning (ML) operators are widely adopted as the building blocks of ML models. 
% 
Among ML operators today, GEneralMatrix Multiplication (GEMM)-based operators are known to be key operators that build the main backbone of ML models.
As their computational overhead dominates the overall execution time (e.g., 42.8\% - 96.6\% in our results), GEMM operators have been the prime optimization targets for fast ML inference.
This led to advanced GPUs and accelerators available today, which provided significant boost in the GEMM performance compared to CPUs, aligned with the lesson from Amdahl's law.
However, accelerating GEMM has significantly shifted the Amdahl's law's landscape for ML inference; due to the decreased GEMM execution time, the relative execution time of non-GEMM operators is now significant.
Although the importance of non-GEMM performance is increasing, we have little knowledge about the non-GEMM performance horizon in the latest hardware platforms and models.
%
%In addition, latest models include more complex non-GEMM operators 
%Nevertheless, the performance of non-GEMM operators has not been studied with the latest GPUs.
%
%In addition, recent models include more complex non-GEMM operators (e.g. GELU), 
%which motivates the characterization of the performance to guide the future ML system design. 

Therefore, to guide non-GEMM-oriented optimizations, we conduct a thorough performance analysis of 17 widely adopted ML models in Hugging Face and Torchvision on workstation and data center platforms with/without GPUs.
We discover that non-GEMM performance bottleneck is a considerable issue across all the platforms and models, accounting for 11.3\% to 73.6\% of total latency, on average. 
The challenge significantly aggravates when we apply quantization, which is a common model compression technique, due to the boosted GEMM performance and extra non-GEMM operators for dequantization and requantization.
To provide insights into non-GEMM optimization targets, we demystify the most dominant non-GEMM operators for each model and deployment software.
We also show that widely adopted optimizations such as operator fusion do not completely address the non-GEMM performance bottleneck, where non-GEMM operators still account for 15\% to 48\% of total latency.
We will open-source our non-GEMM-oriented benchmark framework to facilitate research in non-GEMM optimization.

\end{abstract}

%% file: Sections/01_Introduction.tex
\section{Introduction}
\label{sec:intro}

%Discuss the workload characteristic that led to GEMM acceleration
The success of machine learning (ML) in various problem domains, such as computer vision (CV) ~\cite{he2016deep, he2017mask, sandler2018mobilenetv2, kolesnikov2021ViT, liu2021swin} and natural language processing (NLP)~\cite{kenton2019bert, achiam2023gpt, touvron2023llama}, made ML workloads pervasive in various computing platforms from edge to cloud devices.
ML model inference involves billions of multiply-and-accumulate (MAC) operations (e.g., 497 billions of MAC operations for ResNet 50~\cite{he2016deep}).
Such MAC operations originate from GEneral Matrix Multiplication (GEMM)-based operators, such as CONV2D, Linear, and BMM (batched matrix multiplication).
The GEMM-based operators dominate in terms of the total execution time on CPUs, as shown in~\autoref{fig:Intro_Motivational}.
Therefore, GPUs and accelerators have focused on the optimization of the GEMM-based operators, which significantly enhanced the computational performance (e.g., latency and throughput) of end-to-end ML model inference.

\insertFigure{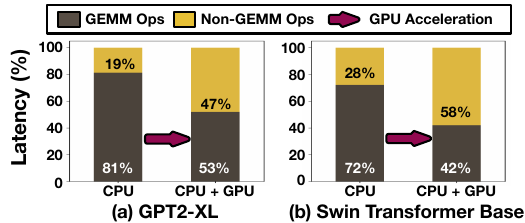}{The latency breakdown into GEMM and non-GEMM operators on AMD EPYC 7763 + NVIDIA A100 GPU. We measure the latency on two popular models from HuggingFace (a) GPT2-XL (batch 1)~\cite{achiam2023gpt} and (b) Swin Transformer (batch 1)~\cite{liu2021swin}.}

% Discuss how the optimizaiton horizon has changed and why we need NonGEMM Bench (Limitation of current practice)
However, because the GEMM operators are being accelerated, the non-GEMM operators, such as memory operations (e.g., reshape, view, and transpose), normalization, and logit computation functions (e.g., Softmax), now account for a considerable amount of the end-to-end latency, compared to that of GEMM operators.
~\autoref{fig:Intro_Motivational} shows the profiled latency breakdown into GEMM and non-GEMM operators running inferences on state-of-the-art large language (GPT2-XL~\cite{achiam2023gpt}) and image classification (Swin Transformer~\cite{liu2021swin}) models. 
The motivational data show that the non-GEMM operations now can account for the majority of the latency with GEMM acceleration, indicating that we now need to consider non-GEMM operators as one of the major optimization targets in the ML system optimization.
However, the research community today lacks a thorough and systematic performance analysis and characterization of non-GEMM operators in the latest models, which hinders the development of non-GEMM oriented optimization techniques.

%no study thoroughly explored the non-GEMM operator performance in latest models, although it is important to quantitatively understand the current status due to the fast evolving GEMM accelerators such as GPUs, and optimization techniques.

% What we did: a characterization 

Therefore, we collect widely-used ML models from Hugging Face~\cite{wolf2020transformers} and Torchvision~\cite{torchvision2016} and perform a thorough performance characterization of of non-GEMM operators in the 17 latest models of four major task domains: Image Classification (IC), Image Segmentation (IS), Object Detection (OD), and Natural Language Processing (NLP).
% 
%Unlike previous work, we analyze non-GEMM operator performance in 18 models collected from Hugging Face~\cite{wolf2019huggingface} and Torchvision~\cite{torchvision2016}, providing important insights on the ML system optimization with powerful accelerators. 
% 
We evaluate the effect of GPU acceleration on the relative latency across GEMM and non-GEMM operators in collected models and conduct deep-dive analysis on the impact of different hardware platforms (workstation and data center), deployment software, and common optimizations (operator fusion and quantization). 
Based on our case studies, we highlight that the non-GEMM performance challenge is common in accelerated inferences and existing optimization techniques (e.g., operator fusion) cannot completely address the challenges.
Also, we demystify the most time-consuming non-GEMM operators in each model, which will help the research community identify non-GEMM operators to be optimized.

To facilitate such research in non-GEMM-oriented optimization techniques, we build an open-source benchmark specialized in non-GEMM performance anlysis, \bench, which will be released after publication.
\bench can profile arbitrary non-GEMM operators supported by PyTorch~\cite{paszke2019pytorch}, ONNX~\cite{onnx}, and TensorRT~\cite{tensorrt}, in addition to the preset of non-GEMM operators collected from the selected 17 popular models, which  provides desired flexibility to users for follow-up research.
%

%We utilize input argument specification extracted from real data to accurately reflect the performance of non-GEMM operators in realistic settings. 

%highlight the contribution of non-GEMM operators to the end-to-end latency of the accelerated model.
%
%We analyze the performance of non-GEMM operators on different hardware configurations ranging from mobile, workstation, to data center, which includes three CPU and three GPU models. 
%
%We also study the impact of common optimization techniques like operator fusion and quantization on the performance of non-GEMM operators. 

% Mention Benchmark
%Finally, 
%To facilitate the characterization of non-GEMM operators in our selected workloads, we also present \bench, an inference benchmarking flow focusing on non-GEMM operators. 
%
%In addition to the preset of non-GEMM operators collected from the selected 18 popular models, \bench can profile arbitrary non-GEMM operators supported by PyTorch~\cite{paszke2019pytorch}, ONNX~\cite{onnx}, and TensorRT~\cite{tensorrt} providing flexibility to users.
% 
%We utilize input argument specification extracted from real data to accurately reflect the performance of non-GEMM operators in realistic settings. 

% List up the contributions

%We will open-source \bench at the time of publication to facilitate the ML system research aligned with the current performance horizon.
%
We summarize our contributions as follows:

\begin{itemize}
    {\item We shed a light on the changed landscape of Amdhal's law in ML system design, which shows the increased importance of non-GEMM operators in systems with GEMM accelerations.}
    
    {\item We perform case studies on two different hardware configurations, workstation and data center, and show the non-GEMM operators are becoming a major consideration across all platforms.}
    
    {\item We identify different dominant non-GEMM operators depending on the model and deployment software flow, which indicates that non-GEMM optimization need to be specialized for each model and deployment software.
    }
    
    {\item We analyze the impact of a common non-GEMM-aware optimization, operator fusion and show that operator fusion does not completely mitigate non-GEMM bottleneck for all models, which motivates follow-up research in non-GEMM performance optimizations}
    
    {\item We evaluate the performance of non-GEMM operators with quantization and quantitatively show the non-GEMM bottleneck aggravates with quantization.}
    
    {\item We open-source \bench, an extensible benchmark flow that enables thorough non-GEMM performance characterization for any model supported by ONNX runtime, TensorRT, and PyTorch, to facilitate non-GEMM-oriented research.}
\end{itemize}

%(1) Reports Non-GEMM operator performances and energy 
%(2) Provides microbenchmark with Non-GEMM operators with input parameters based on real inputs from dataset (no synthetic data)
%(3) We covers most popular 16 Hugging face and Torch Vision models in image classification, object detection, segmentation, language models 
%(4) Thorough case studies on seven different hardware configurations on ARM (Apple M2 Max) and X86 (Intel and AMD) CPUs and three different NVIDIA GPUs.

%CPU-only
%- M2
%- i7-....m
%- EPYC
%- i9-13900k

%CPU + gPU
%- i7-....m + RTX 4060m
%- EPYC + A100
%- i9-13900k + RTX 4090

%Verified to run on mobile, desktop, and server 

%% file: Sections/02_Background.tex
\section{Background}
\label{sec:background}

% \insertWideFigureShrink{Background_Operator}{\TODO{Fix Labeling}A description of Linear (a) and Conv1D (c) operators as GEMM operators example,  and that of non-maximum suppression~\cite{he2017mask} (b) and Layer Normalization~\cite{ba2016layernorm} (d) as a non-GEMM operator example.}{0.75}

\insertWideFigureShrink{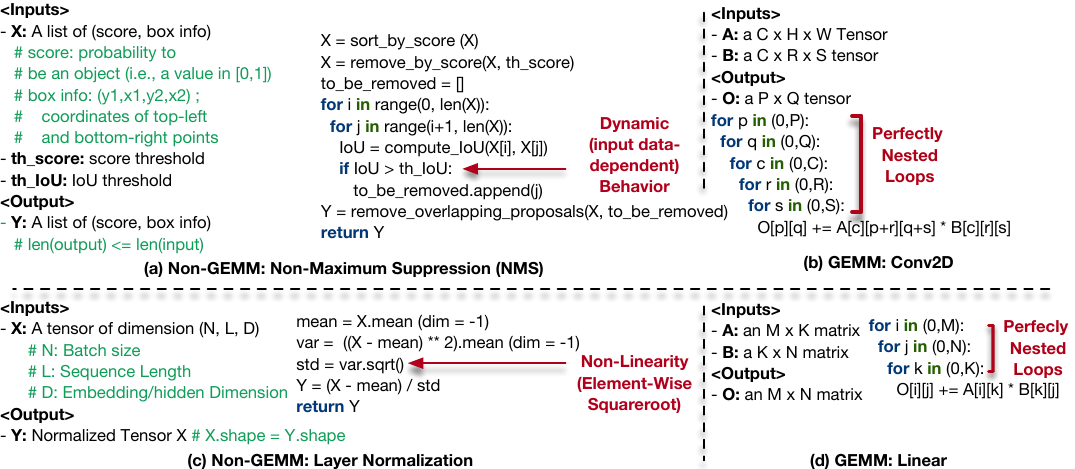}{Descriptions of example non-GEMM and GEMM operators. (a) (Non-GEMM) non-maximum suppression~\cite{he2017mask}, (b) (GEMM) Conv1D (c) (Non-GEMM) Layer Normalization~\cite{ba2016layernorm}, and (d) (GEMM) Linear.}{1}

%We discuss the ML operators and recent ML models.

\subsection{ML Operators}
\label{subsec:background_operators}

ML operators are the building blocks of ML models, which define the computation over one or multiple input tensors. 
% 
%Operators consist of a computation or operation that generates one or multiple output tensors from input tensors. 
Examples include convolution (Conv2d), matrix multiplication (linear, BMM, etc.), activation, and normalization, as listed in~\autoref{fig:Background_Operator}. 
We categorize operators into two classes: GEMM operation-based ("GEMM operators") and the others ("Non-GEMM operators").
We discuss each class of ML operators next.

\betterparagraph{GEMM-based Operators (GEMM Operators)}
%\label{subsubsec:gemm_ops}
%
%% Introduction to GEMM-based operators
GEMM-based operators (or \textit{GEMM operators}) refer to all the ML operators that can be represented as a matrix multiplication operation, which include linear, Conv2d, and batched-matrix multiplication (BMM).
For example, ~\autoref{fig:Background_Operator} (b) and (d) illustrate two popular GEMM operators: Linear and Conv2d operators, respectively. 
Each operator can be represented into a perfectly nested loop with multiply-and-accumulate (MAC) operation in the inner-most loop.
Note that variants that are not matrix multiplication in the default form like Conv2d can be converted into GEMM (e.g., im2col~\cite{chetlur2014cudnn}), which motivated the term, GEMM operator.

%loop over the input tensor dimensions and populate the output tensor by computing a MAC inside the inner most loop.
%
%Despite having different semantics, GEMM operators compute MACs in tightly nested loops and can be expressed as a series of matrix multiplications or dot products. 
% 
%For example, 2-dimensional convolution can be mapped to a matrix multiplication as shown in ~\cite{chetlur2014cudnn}, and linear layers compute a vector-matrix multiplication between an input feature tensors and the layers' weights. 
% 

%% Properties of GEMM-based operators
GEMM-based operators are known to be compute-intensive, which accounts for the majority of the execution unless accelerated by GPUs or accelerators, as CPU results in~\autoref{fig:Intro_Motivational} show.
However, they have regular computation patterns that can be summarized as a perfectly nested for loop.
The regular pattern allows various loop optimization techniques such as loop reordering, tiling, and parallelization, which is referred to as dataflow~\cite{parashar2019timeloop, kwon2019understanding, yang2020interstellar}
With the dominance of GEMM operators in execution time and high optimization potential together, GEMM operators have been the prime optimization target for acceleration, which led to high-performance GPUs (e.g., H100~\cite{nvidia_h100}) and accelerators~\cite{jouppi2017tpu }.

\betterparagraph{Non-GEMM Operators}
%\label{subsec:nongemm_ops}
%
%% Introduction to non-GEMM operators
Non-GEMM operators refer to all ML operators other than GEMM operators.
They span various functionalities (e.g., memory layout manipulation and normalization) other than applying weights to input tensors.
Because of their diverse functionalities, their computation patterns are often not a perfectly nested loop with MAC, which can also involve non-linear functions and memory-oriented operations.
% 
%They encompass linear and non-linear functions, as well as memory operations. 
For example, ~\autoref{fig:Background_Operator} (a) shows non-maximum suppression (NMS) operator often found in R-CNN model variants~\cite{he2017mask, ren2015faster}.
As found in the example, the entire operation cannot be summarized into single perfectly-nested loop, which involve other operations such as sort and filtering.
In addition, the operation involves a conditional statement, which introduces non-deterministic behaviors to the operator.
The layer normalization example in~\autoref{fig:Background_Operator} (c) also shows another key characteristic of the non-GEMM operators: non-linear functions.
Because of such characteristics distinguished from GEMM operators, optimization methodologies for GEMM operators cannot be applied to accelerate non-GEMM operators.

To understand the extent of the non-GEMM operators, we analyze non-GEMM operators in 17 recent models in the computer vision and natural language processing domains.
We select models based on their popularity in the Hugging Face to obtain realistic workload.
We list the models we investigated in~\autoref{tab:modeltable}.
Based on our analysis, we categorize non-GEMM operators based on their functionality and summarize their usage in models and characteristics in~\autoref{tab:microbench}. 
% 
%

%Operators Groups 
\begin{itemize}
\label{nongemm_groups}
{\item{\textbf{Normalization.} Normalization operators regularize the data range across a selected dimension using the mean and standard deviation.
Examples include BatchNorm~\cite{ioffe2015batchnorm} and LayerNorm~\cite {ba2016layernorm}, which are widely adopted in computer vision and NLP models~\cite{he2017mask, achiam2023gpt}.
%
%Such variants defer by the choice of the normalized dimension (e.g., BatchNorm normailizes a tensor across the batch dimension).
%For a tensor with dimensions (\textbf{B}atch,~\textbf{C}hannel,~\textbf{H}eight,~\textbf{W}idth), Batch Norm normalizes over the batch dimension B while Layer Norm performs the normalization across the last dimension of the tensor (\autoref{fig:Background_Operator} (d)).
%
RMS Norm~\cite{zhang2019root}, which is adopted in recent large language models~\cite{touvron2023llama}, is another example of the normalization function.
RMS Norm eliminates the division by standard deviation in typical normalization functions and performs $\sqrt{\frac{1}{n}\Sigma^{n}_{i=1} (X_{i} - \mu)}$, where $X_{i}$, n, and $\mu$ refer to the i-th data, number of data, and the mean, respectively.
}
}
{\item{\textbf{Activation.} Activation operators introduce non-linearity into the model.
Rectified Linear Unit (ReLU) function~\cite{nair2010relu} is an example of activation operators widely used in CNN based ML models ~\cite{sandler2018mobilenetv2, he2016deep, simonyan2014very}. 
ReLU injects non-linearity into the model based on the sign of the data by applying the element-wise function, $ReLU(X) = Max(0,X)$.
Another variant of activation operators is the the Gaussian Error Linear Units function (GELU)~\cite{hendrycks2016gaussian}, which is a popular activation function adopted in transformer based ML models~\cite{achiam2023gpt, xie2021segformer, kolesnikov2021ViT, liu2021swin}.
Unlike ReLU simply gates out negative values to be 0, GELU requires to compute the Cumulative Distribution Function (CDF) of a Gaussian distribution, which is often denoted as $\phi$. %accounts for the value of the data when inserting the non-linearity and not only the sign. 
GELU multiplies the input $X$ by the Cumulative Distribution Function (CDF) of a Gaussian distribution ($\phi$): $GELU (X) = X * \phi(X)$~\cite{hendrycks2016gaussian}.
}}
{\item{\textbf{Memory Operators.} Memory operators are responsible for the memory allocation and the layout modification of tensors. 
%
%Examples include the reshape and concatenation (cat) operators.
% 
For example, Reshape modifies the shape (e.g. dimension order) of a tensor and return a new view of the tensor following the new dimension order. 
%
%While reshape operators do not allocate additional memory, the cat operator allocates new memory to store a newly created tensors by concatenating the input tensors along a new dimension.
% 
% An example of memory operations would be the Transpose operator used in the implementation of the self-attention block of language models shown in~\autoref{fig:models_architecture} (c). 
% An example of memory operations would be the PyTorch reshape function used in the implementation of the attention block in Hugging Face transformers library to ensure the correct dimensions of the Key matrix before multiplying it with the Query matrix.
%
}}
%
% {\item{\textbf{Element-Wise Arithmetic.} Element-wise arithmetic operators cover arithmetic operators applied on tensors in an element . 
%
{\item{\textbf{Element-Wise Arithmetic.} Element-wise arithmetic operators refer to all the operations applied on individual elements in a tensor (other than activations).
For example, \autoref{fig:models_architecture} (c) contains an element-wise division applied to scale the elements of tensors in the attention block}.}
{\item{\textbf{RoI Selection.} RoI selection operators are found in R-CNN variants.~\cite{ren2015faster,he2017mask}. 
They filter down bounding boxes proposed by the region proposal network (\autoref{fig:models_architecture} (b)) and align the remaining boxes to the objects detected in the image.
Non-Maximum Suppression (NMS) is an example of RoI Selection, which is described in \autoref{fig:Background_Operator}. 
Given a list of scores and bounding box information, it selects bounding boxes by applying the Intersection over Union (IoU) metric.
}}
\end{itemize}

\insertFigure{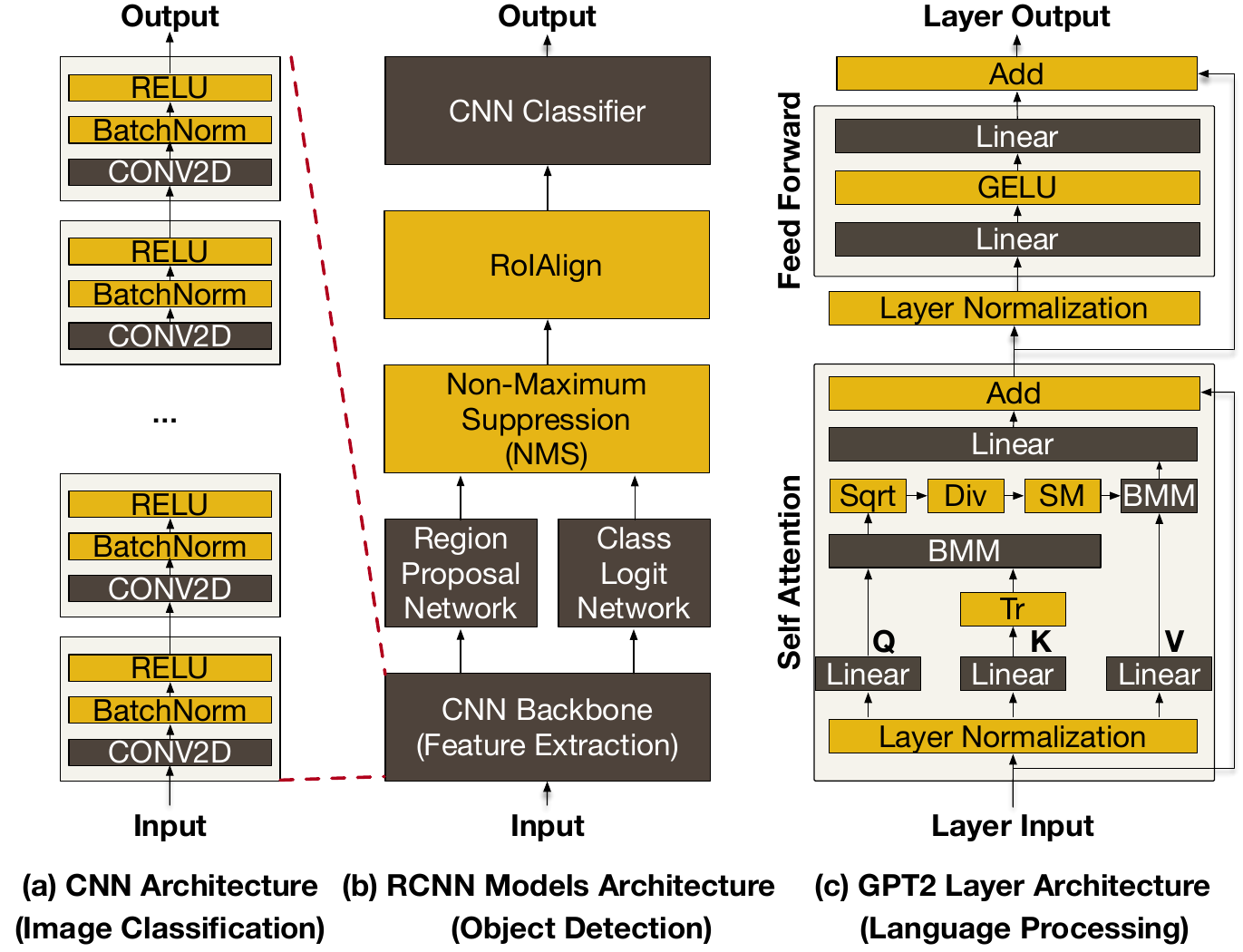}{Architectures of three popular ML model families. \vspace{-4mm}}

\input{Tables/tab_microbench_new}
\subsection{ML Models and Popular Tasks}
\label{subsec:models_and_popular_tasks}

The heterogeneity in non-GEMM operators enabled ML developers to build models supporting a wide range of modalities and tasks (e.g. computer vision and NLP). 
As highlighted in ~\autoref{fig:models_architecture}, computer vision (\autoref{fig:models_architecture} (a) and (b)) and NLP (\autoref{fig:models_architecture} (c)) models are characterized by distinct architectures leveraging unique combinations of GEMM and non-GEMM operators. 

For example, traditional image classification models are often based on the convolutional neural network (CNN) architecture, which cascades GEMM (Conv2d) and non-GEMM (normalization and activation)operators~\cite{sandler2018mobilenetv2, he2016deep}.  
Object detection models, such as Mask R-CNN~\cite{he2017mask}, often utilize CNNs for feature extraction, region proposal, and classification, as illustrated in~\autoref{fig:models_architecture} (b). Unlike image classification models, they combine the CNNs with unique non-GEMM operators such as non-maximum suppression (NMS) and ROI Align to process and filter the bounding boxes for objects.
On the other hand, recent language models employ the transformer architecture, which leverages the attention mechanism introduced in~\cite{vaswani2017attention}.
Transformers combine a unique set of GEMM (BMMs and Linear) and non-GEMM (normalization, memory, and element-wise arithmetic) operators, as shown in~\autoref {fig:models_architecture} (c).  

As we can find in the aforementioned examples, the model architectures and the combination of GEMM and non-GEMM operators are diverse. This would mean that the performance implication of non-GEMM operators would vary across models, as our motivational data presented in ~\autoref{fig:Intro_Motivational} show. 
This motivates a thorough characterization study that investigates (1) if the non-GEMM performance challenge is pervasive across popular models and (2) how significant their implication is, under widely adopted optimization techniques (e.g., integer quantization~\cite{jacob2018quantization} and operator fusion~\cite{niu2021dnnfusion}).  
Therefore, we conduct a thorough case study of the non-GEMM performance horizon.
%
%We discuss our methodology of the workload characterization study and present the results next.

%Such a diversity in non-GEMM operators and model architectures across tasks paired the preliminary data presented in~\autoref{fig:Intro_Motivational} motivate us to study the impact of non-GEMM operators have on the inference performance of popular models.

~\insertWideFigure{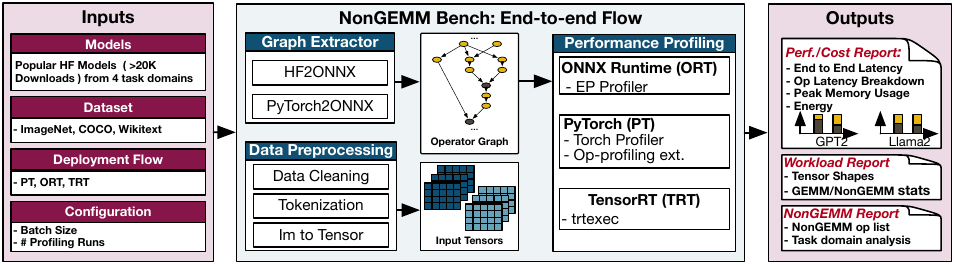} {An overview of \bench flow. }

% Paired with the motivational data presented in ~\autoref{fig:Intro_Motivational}, the diversity in non-GEMM operators motivates us to study the impact of imploof have diverse architectures to support multiple task domains. 
% % 
%
%In \bench, we collect models from the computer vision and NLP domains to breakdown their performance and understand the contributions of each operator, especially the non-GEMM ones. 

%State of the art ML models often interleave non-GEMM operators in various ways among GEMM operators to support a broad range of applications. 
%
%This practice leads to increased non-GEMM operator heterogeneity in model architectures across application domains as illustrated in ~\autoref{fig:models_architecture}.
% 
%As a result, this operator heterogeneity uniquely affects the inference performance of each model.
%
%Therefore, we construct \bench to better understand the performance landscape of popular state of the art ML models.
% 

%% file: Tables/tab_microbench_new.tex
\begin{table*}
\centering
\caption{Non-GEMM operators in eight selected model variants from~\autoref{tab:modeltable} and their characteristics. Example input shapes are captured based on inferences using real datasets.}
\scriptsize

\begin{tabular} {| c | c |  c | c | c |c| c | c| c|}
\hline 

\multirow{2}{*}{
\begin{tabular}[c]{@{}c@{}} 
\textbf{Operator} \\ \textbf{Group}
\end{tabular}}

% \multicolumn{1}{|c|}{\textbf{Operator Group}} 

& \multirow{2}{*}{
\begin{tabular}[c]{@{}c@{}} 
\textbf{Operator}
\end{tabular}}

% & \multicolumn{1}{c|}{
% \textbf{Operator}} 

& \multirow{2}{*}{
\begin{tabular}[c]{@{}c@{}} 
\textbf{Model}
\end{tabular}}

% & \multicolumn{1}{c|}{\textbf{Model}}

& \multirow{2}{*}{
\begin{tabular}[c]{@{}c@{}} 
\textbf{Single} \\ \textbf{Operation}
\end{tabular}}

% & \multicolumn{1}{c|}{\textbf{Single Operation}}

& \multirow{2}{*}{
\begin{tabular}[c]{@{}c@{}} 
\textbf{Single} \\ \textbf{Operand}
\end{tabular}}

% & \multicolumn{1}{c|}{\textbf{Single Operand}}

& \multirow{2}{*}{
\begin{tabular}[c]{@{}c@{}} 
\textbf{Non} \\ \textbf{Linearity}
\end{tabular}}
% & \multicolumn{1}{c|}{\textbf{Non-Linearity}}

& \multirow{2}{*}{
\begin{tabular}[c]{@{}c@{}} 
\textbf{Dynamicity}
\end{tabular}}
% & \multicolumn{1}{c|}{\textbf{Dynamicity}}

& \multirow{2}{*}{
\begin{tabular}[c]{@{}c@{}} 
\textbf{Reduction}
\end{tabular}}
% & \multicolumn{1}{c|}{\textbf{Reduction}}

& \multirow{2}{*}{
\begin{tabular}[c]{@{}c@{}} 
\textbf{Example} \\ \textbf{Input Shape}
\end{tabular}}
\\
&&&&&&&& \\
% & \multicolumn{1}{c|}{\textbf{\# Input Shapes}} \\

\hline 

\multirow{4}{*}{Activation}
& ReLu 
& DETR
& \checkmark
& \checkmark
&  %checkmark
&  %\checkmark
&  %\checkmark
%Torch.nn.modules.activation
& [2,64,533]\\%\tiny [2,64,533],  \\
\cline{2-9} 

& GELU
& ViT-l16
%Torch.nn.modules.activation
& 
& \checkmark
& \checkmark
& 
& 
& [1, 97, 4096] \\%\tiny [1, 97, 4096] \\
\cline{2-9}

& GELU 
& GPT2-XL 
%transformers.activations.GELUActivation
& 
& \checkmark
& \checkmark
&  
&  
& [1, 8, 6400] \\%\tiny [1, 8, 6400] \\
\cline{2-9}

& SiLu 
& Llama-2 
& 
& \checkmark
& \checkmark
&  
& 
& [1, 10, 11008] \\  %\tiny [1, 10, 11008]\\%, [1, 66, 11008], [1,338,11008] \\
\hline

\multirow{6}{*}{Normalization}
& LayerNorm
& Segformer
&  %Torch.nn.modules.normalization
& \checkmark
& \checkmark
&  
& \checkmark
& [2, 16384, 32] \\%\tiny [2, 16384, 32]\\%, [2, 256, 32], [2, 4096, 64] \\
\cline{2-9} 

& BatchNorm2d
& Segformer
&   %Torch.nn.modules.batchnorm
& \checkmark
& \checkmark
&  
& \checkmark
& [2, 256, 128, 128] \\%\tiny [2, 256, 128, 128]\\% \\
\cline{2-9} 

& LlamaRMSNorm
& Llama
&  %transformers.models.llama
& \checkmark
& \checkmark
&  
& \checkmark
& [1, 10, 4096] \\%\tiny [1, 10, 4096]\\%, [1,338,4096], [1,118,4096]    \\ 
\cline{2-9} 

& FrozenBatchNorm2d
& MaskRCNN
&   %torchvision.ops.misc
& \checkmark
& \checkmark
&  
& \checkmark
& [1, 1024, 50,68] \\%\tiny [1, 1024, 50,68]\\%, [1, 256, 100, 136] \\
\cline{2-9} 

& FrozenBatchNorm2d
& DETR
&   %transformers.models.detr
& \checkmark
& \checkmark
&  
& \checkmark
& [1, 2048, 25, 34] \\%\tiny [2, 850, 256] \\
\cline{2-9} 

& LayerNorm
& DETR
&   %Torch.nn.modules.normalization
& \checkmark
& \checkmark
&  
& \checkmark

& [2, 850, 256] \\%\tiny [2, 850, 256]\\%, [2, 100, 256] \\
\hline 

\multirow{5}{*}{ Elmt-wise Arithmetic} 
& Add
& Segformer
& \checkmark %Torch.add
&  
&  
&  
& 
& [2, 16384, 32] \\%\tiny [2, 16384, 32]\\%, [2, 4096, 64], [2, 256, 256] \\
\cline{2-9} 

& Mul
& Llama-2
& \checkmark  %Torch.mul
&  
&  
&  
&  
& [1, 10, 11008] \\%\tiny [1, 10, 11008]\\%, [1, 66, 11008] \\ 
\cline{2-9} 

& Neg
& Llama-2
& \checkmark %Torch.neg
&  
&  
&  
&  
& [1, 32, 10, 64] \\%\tiny [1, 32, 10, 64] \\
\cline{2-9}

& TrueDiv
& Segformer 

& \checkmark %Torch.true\_divide
&  
&  
&  
&  
& [2, 1, 16384, 256] \\%\tiny [2, 1, 16384, 256]\\%, [2, 8, 256, 256] \\
\cline{2-9} 

& TrueDiv 
& GPT2-XL
& \checkmark %Torch.true\_divide
&  
&  
&  
&  
& [1, 25, 8, 8] \\%\tiny [1, 25, 8, 8] \\
\hline 

% contiguous & Memory Op & Returns a new allocated tensor in a specified memory format \cite{paszke2019pytorch} & [391, 144, 4, 32] \\ 
% \hline 

% permute & Memory Op &  Returns a different view of the tensor based on the specified dimensions ~\cite{paszke2019pytorch} & [782, 144, 4, 32] \\ 
% \hline 

% view & Memory Op  & Allocates a new tensor in memory with same data as the input but with different shapes & [391, 12, 12, 1] \\ 
% \hline 

\multirow{9}{*}{Memory}  
& Contiguous
& Segformer
& \checkmark %Torch.Tensor.contiguous
& \checkmark
&  
&  
&  
& [2, 32, 128, 128] \\%\tiny [2, 32, 128, 128]\\%, [2, 4096, 2, 32], [2, 64, 64, 64] \\
\cline{2-9} 

& Contiguous
& Llama-2
& \checkmark %Torch.Tensor
& \checkmark
&  
&  
&  
& [1, 10, 32, 128] \\%\tiny [1, 10, 32, 128]\\%, [1, 118, 32, 128] \\
\cline{2-9} 

& Permute
& ViT-b16
& \checkmark %Torch.permute
& \checkmark
&  
& 
&  
& [1, 768, 196] \\%\tiny [1, 768, 196]\\% \\ 
\cline{2-9} 

& Permute
& GPT2-XL
& \checkmark %Torch.permute
& \checkmark
&  
&  
&  
& [1, 8, 25, 64] \\%\tiny [1, 8, 25, 64] \\
\cline{2-9} 

& Split
& GPT2-XL
& \checkmark %Torch.split
& \checkmark
&  
&  
&  
& [1, 8, 4800] \\%\tiny [1, 8, 4800] \\ 
\cline{2-9} 

& View 
& GPT2-XL
& \checkmark %Torch.Tensor.view
& \checkmark
&  
&  
&  
& [1, 8, 1600] \\%\tiny [1, 8, 1600]\\
\cline{2-9} 

& Reshape
& ViT-b16
& \checkmark %Torch.reshape
& \checkmark
&  
&  
&  
&[1, 768, 14, 14] \\%\tiny [1, 768, 14, 14]\\
\cline{2-9} 

& Expand
& ViT-b16
& \checkmark %Torch.Tensor.expand
& \checkmark
&  
&  
&  
& [1, 1, 768] \\%\tiny [1, 1, 768] \\
\cline{2-9} 

& Squeeze
& Llama-2
& \checkmark %Torch.squeeze
& \checkmark
&  
&  
&  
& [1, 1, 10, 128] \\%\tiny [1, 1, 10, 128] \\ 
\hline 

%\multirow{2}{*}{Logit Computation} 
Logit 
& Softmax 
& DETR 
& \checkmark %Torch.nn.Functional.softmax
& \checkmark
& \checkmark
&  
& \checkmark
& [1, 25, 8, 8] \\%\tiny [1, 25, 8, 8] \\ 
\cline{2-9} 

Computation
& Softmax
& Segformer
& \checkmark %Torch.nn.Functional.softmax
& \checkmark
& \checkmark
& 
& \checkmark
& [2, 1, 16384, 256] \\%\tiny [2, 1, 16384, 256]\\%, [2, 5, 1024, 256] \\ 
\hline 

RoI Selection
& NMS
& MaskRCNN
&  %torchvision.ops.nms
& 
& 
& \checkmark
& 
& [4663, 4] \\%\tiny [4663, 4] \\ 
\hline 

Interpolation
& Interpolate
& Segformer
&  %
& \checkmark
& 
& 
& 
& [2, 256, 128, 128] \\%Torch.nn.Functional
%, [2, 256, 16, 16] \\
\hline 

\end {tabular}
\vspace{-4mm}
\label{tab:microbench}
\end{table*}

%% file: Sections/03_Benchmark.tex
\vspace{-4mm}
\section{Performance Characterization Methodology}
\label{sec:bench}

% ~\insertWideFigure{bench_overview_jun} {Overview of the different stages of \bench \HK{This figure could be squeezed more. Try to reduce the height of the figure.}} 

% List the requirements to a good ML benchmark focusing on non-GEMM operators
% Requirement 1) Needs cover diverse task domains
% Requirement 2) Need to be based on frequently used models
% Requirement 3) Need to support broad task domain
% ...

% (High-level Flow: Discuss how NonGEMM Bench is designed to satisfy the requirements)

%% non-GEMM op benchmark requirements
To understand the realistic performance landscape of the latest ML models with non-GEMM workloads, we must (1) capture operator level performances in end-to-end inferences, (2) use widely-used models by the research community and industry, (3) cover diverse task domains, and (4) use real datasets. 
%% Discuss existing benchmarks fail to satisfy all reqruiements
However, ML Benchmarks available today (e.g., MLPerf~\cite{reddi2020mlperf}), unfortunately, do not satisfy all the requirements since they do not focus on the non-GEMM operators.
Long-tail bench~\cite{longtailbench} identified a similar problem as this work, but it focuses on a limited set of custom kernels, which fails to represent broad task domains.
Therefore, to facilitate our non-GEMM operators analysis and better-understand the impact of non-GEMM operators on system performance, we develop a new ML benchmark, \bench. 
\bench provides operator-level breakdown of end-to-end inference latency in the operator graph level, which enables detailed non-GEMM operator performance analysis, as we present in~\autoref{sec:case_studies}. 
% 
%In addition, to facilitate non-GEMM operator-focused system optimizations, \bench includes a micro-benchmark of non-GEMM operators with concrete input arguments and tensors collected from realistic datasets and models.
%
To capture the performance in the latest ML workload, we select 17 highly downloaded (more than 10K downloads on average) models from HuggingFace~\cite{hfmodels} to enhance the \textit{representativeness} of \bench and our analysis. 
We discuss the models and datasets adopted in \bench in detail and describe the structure of \bench next.

%Previous state-of-the-art work ~\cite{longtailbench} focused on micro-benchmarking operators without device specific kernel implementations. 
%

% \subsection{\bench Models}
\subsection{Models included in \bench}
\label{sub:model_description}
% Discuss the choice of workload, Rerpresentativeness of NonGEMM Bench

~\autoref{tab:modeltable} lists the \bench model registry which contains 17 models based on state-of-the-art CNN and Transformer architectures with number of parameters ranging from 3.7M to 7B, demonstrating the \textit{diverse} model coverage of \bench.
The selected models cover four major task domains in ML, which include Image Classification (IC), Object Detection (OD), Image Segmentation (IS), and Natural Language Processing (NLP).
% 

%\subsubsection{\textbf{Computer Vision (CV) Domain}}
%\label{subsub:cv}

%Deep learning is widely used in the CV domain~\cite{kolesnikov2021ViT, he2016deep,ren2015faster}.
%\bench covers three important tasks in the CV : image classification (IC), image segmentation (IS), and object detection (OD)
% 

\betterparagraph{Image Classification (IC)} Image classification refers to a CV task that identifies a class label of a given image. Image classification models extract features (i.e., high-level and dimensional information of the input image) from an input image and report the class label utilizing the features. \bench includes six most popular IC models in HuggingFace~\cite{hfmodels}: Three variants of Vision Transformer~\cite{kolesnikov2021ViT}, and three variants of Swin Transformer ~\cite{liu2021swin}.

\betterparagraph{Image Segmentation (IS)} Image segmentation refers to a computer vision task that identifies the area in a image for each class. Like IC models, IS models also extract features and utilize them to identify objects located in an image and spatially separate them by highlighting pixels that belong to each object.
\bench includes two state-of-the-art IS models: Segformer ~\cite {xie2021segformer} and MaskFormer ~\cite{cheng2021per}.

\betterparagraph{Object Detection (OD)} Object detection refers to a computer vision task that identifies the location of objects in an image and outputs the bounding box of each object. OD models extract features and generate region proposals, which refer to the candidate locations and bounding boxes of objects. Using Region of Interest (RoI) processing non-GEMM operators in~\autoref{tab:microbench}, OD models refine raw region proposals generated by a region proposal network. The refined RoIs are used as inputs to the CNN classifier at the end, and the classifier determines the class label of objects inside each refined RoI. \bench includes three popular OD models~\cite{hfdetection}: FasterRCNN ~\cite{ren2015faster}, MaskRCNN ~\cite{he2017mask}, and DETR ~\cite{carion2020end}.

\betterparagraph{Natural Language Processing (NLP)}
%\label{subsub:nlp}
% %
NLP refers to tasks involving the analysis and understanding of human (natural) language.
% %
NLP models extract context and features from an input text sequence and use the extracted context and features to perform multiple applications like translation and text generation~\cite{zhu2020incorporating}~\cite{brown2020language}.
Transformer \cite{vaswani2017attention} based DNNs have become the dominant model architecture in NLP and are the backbone of popular state-of-the-art generative large language models like GPT ~\cite{radford2019language} and Llama ~\cite{touvron2023llama}. 
~\autoref{fig:models_architecture} (c) shows the layer architecture of GPT's transformer. It consists of a self-attention block built by cascading GEMM operators with Normalization, Memory, Logit Computation and Element-wise Arithmetic non-GEMM operators (~\autoref{tab:microbench}). 
\bench includes six popular NLP models~\cite{hfnlp}: Bert~\cite{kenton2019bert}, three variants of GPT2~\cite{achiam2023gpt}, Llama2-7b~\cite{touvron2023llama}, and Mixtral 8x7B~\cite{jiang2024mixtral}.

%To enable the thorough analysis of the landscape of non-GEMM performance presented in \autoref{sec:case_studies}, \bench provides an end-to-end profiling flow to breakdown the inference latency at the operator level. 
%
%We first discuss the \bench's flow with its input and output.

%\subsection{\bench Inputs}
%\label{sub:benc_inputs}

%We discuss the input and output of \bench and describe the components in the benchmark.

% 
% \bench provides two flows: end-to-end inference and microbench flows.
% %
% The End-to-end flow evaluates the performance of operators when the model is running inference on dataset data. 
% %
% The Microbench flow performs micro-benchmarking on the operator by running it in a standalone fashion using synthetic data with realistic dimensions. 

\subsection{\bench Inputs}
\label{subsec:bench_in}

As shown in~\autoref{fig:bench_overview_shrinked},  \bench flow receives workload and dataset information (default: 17 \bench models in~\autoref{tab:modeltable}), target deployment flow, and other configurations such as the batch size and number of runs for performance characterization.
%
%We describe each input in detail as follows.
%
%target deployment software flow (PyTorch, ONNX runtime, or TensorRT), dataset, and other configurations listed in~\autoref{fig:bench_overview_shrinked} as inputs.
%
%Optionally, users can provide custom models with associated datasets to profile their workloads.
%
%By default, \bench runs the 17 models in~\autoref{tab:modeltable}.
%Note that models are optional; if target models are not specified by a user, \bench will run default 15 models included in the benchmark.
%

\betterparagraph{Models} As described in ~\autoref{sub:model_description}, \bench includes a registry of 17 selected popular ML models. 
Nonetheless, we designed \bench to be easily expandable to accommodate rapidly evolving ML models that constantly introduce new operators.
Users can benefit from the features of our benchmark by simply plugging their new models into the \bench model registry (\autoref{fig:bench_overview_shrinked}) by specifying the model class, its weights and any data preprocessor.

\betterparagraph{Deployment Flow} \bench supports four popular inference frameworks: ONNX Runtime~\cite{onnxruntime}, PyTorch~\cite{paszke2019pytorch},  TensorRT~\cite{tensorrt}, and TorchInductor~\cite{ansel2024pytorch2}. 

\betterparagraph{Datasets} To evaluate the models, \bench utilizes real datasets popular in each domain. 
We use ImageNet 2012 ~\cite{imagenet_cvpr09} and MS COCO ~\cite{lin2014microsoft} for computer vision tasks.
As for language models, we use wikitext dataset \cite{wikitext} available on HuggingFace.
For custom models, \bench allows users to specify their own dataset for their models.

\betterparagraph{Configurations} Users can specify detailed configurations for the performance characterization using \bench. The configurations include the batch size, the number of profiling iterations, and the target device.

\subsection{\bench Outputs}
\label{subsec:bench_out}

\bench generates many statistics organized into three categories: performance, workload, and non-GEMM-specific reports.

\betterparagraph{Performance Report}
The performance report includes key performance metrics such as the end-to-end latency with operator level break-downs (\autoref{fig:server_op_breakdown_camera_ready}) and the end-to-end energy consumption (\autoref{fig:server_torch_energy_breakdown_Transposed}).

\betterparagraph{Workload Report}
The workload report includes the types of operators and the shape of the tensors for each operator captured during inference on realistic data.

\betterparagraph{Non-GEMM Report}
The non-GEMM report provides insights on non-GEMM operators, such as the number of operator variants of the same class of non-GEMM operator (e.g., DETR~\cite{carion2020end} employs two variant of BatchNorm, a custom implementation and one in the PyTorch operator library) and non-GEMM operator trace on different domains.

\input{Tables/tab_model_table_fixed}

\subsection{\bench Performance Characterization Flow}
\label{e2eflow}
%The end-to-end inference flow profiles \bench models running inferences on data samples from the selected dataset.
\bench's software flow accepts inputs described in~\autoref{subsec:bench_in} and generates outputs described in~\autoref{subsec:bench_out}.
Internally, \bench includes graph extractor, data preprocessing, and performance profiling modules.

\betterparagraph{Graph Extractor} The \textit{Graph Extractor} module extracts the operator graphs of input models based on the selected deployment flow. 
\bench utilizes graph exporters in the HuggingFace Transformers~\cite{wolf2020transformers} and PyTorch.

\betterparagraph{Data Preprocessing} The \textit{Data Preprocessing} module includes model-specific preprocessing functions that fetch raw data from the target dataset, clean the data, and apply desired transformations (e.g., tokenization and image to tensor). % to generate input tensors for the profiling stage. 

\betterparagraph{Performance Profiling}
The \textit{Performance Profiling} (PP) module launches the inferences, collects performance statistics, and generates output reports discussed in~\autoref{subsec:bench_out}.
The module selects appropriate profiling functions based on the deployment flow choice.
For PyTorch, the PP module utilizes the PyTorch Profiler \cite{pytorch_profiler}.
For TensorRT~\cite{trt_oss}, the PP module leverages its profiling APIs.
For ONNX Runtime~\cite{onnxruntime}, the PP  module invokes Execution Provider profiling.

%% file: Tables/tab_model_table_fixed.tex
\begin{table}

% \caption{Tasks and Models in \bench}
\caption {Tasks and Models Evaluated in \autoref{sec:case_studies}}
\centering

\scriptsize
\begin{tabular} { |c|c|c|c |} 
\hline
%
% \multirow{2}{*}{
% \begin{tabular}[c]{@{}c@{}} 
% Domain  \\  
% \end{tabular}} 
% & 

%
% \textbf{ Domain }  &
\textbf{ Application }  & \textbf{ Models }  & \textbf{ \# Params }  & \textbf{ Dataset } \\
\hline

% \multirow{13}{*}{
% \begin{tabular}[c]{@{}c@{}} 
% Computer  \\ Vision 
% \end{tabular}
% } &
\multirow{7}{*}{
\begin{tabular}[c]{@{}c@{}} 
Image \\ Classification (IC)
\end{tabular} 
}
& ViT base (Vt-b)~\cite{kolesnikov2021ViT}  & 307M 
& \multirow{7}{*}{
\begin{tabular}[c]{@{}c@{}} 
ImageNet \\ ~\cite{imagenet_cvpr09}
\end{tabular} 
}
\\ 
\cline{2-3}
& ViT large (Vt-l)~\cite{kolesnikov2021ViT}  & 307M 
&
\\ 
\cline{2-3}
& ViT huge (Vt-h)~\cite{kolesnikov2021ViT}  & 632M &  \\ 
\cline{2-3}
& Swin tiny (Sw-t)~\cite{liu2021swin} & 29M &  \\ 
\cline{2-3}
& Swin small (Sw-s)~\cite{liu2021swin} & 50M &  \\ 
\cline{2-3}
& Swin base (Sw-b)~\cite{liu2021swin} & 88M &  \\ 
\hline
 \multirow{3}{*}{
\begin{tabular}[c]{@{}c@{}} 
Object \\ Detection (OD)
\end{tabular} 
} & FasterRCNN (FRCNN)~\cite{ren2015faster} & 42M & \multirow{3}{*}
{
\begin{tabular}[c]{@{}c@{}} 
COCO \\~\cite{lin2014microsoft}
\end{tabular} 
} \\
\cline{2-3}
& MaskRCNN (MRCNN)~\cite{he2017mask} & 44M &  \\
\cline{2-3}
& DETR~\cite{carion2020end} & 41M & \\ 
\hline
\multirow{2}{*}{
\begin{tabular}[c]{@{}c@{}} 
Image \\ Segmentation (IS)
\end{tabular}} & Maskformer (MF)~\cite{cheng2021per} & 102M & \multirow{2}{*}{
\begin{tabular}[c]{@{}c@{}} 
COCO \\~\cite{lin2014microsoft}
\end{tabular} 
}\\
\cline{2-3}
& SegFormer (Seg)~\cite{xie2021segformer} & 3.7M &  \\
\hline
% 
% \multirow{5}{*}{NLP} 
 \multirow{5}{*}{
\begin{tabular}[c]{@{}c@{}} 
Natural Language \\ Processing (NLP)
\end{tabular} 
} 
& GPT2~\cite{achiam2023gpt} & 117M & \multirow{5}{*}{
\begin{tabular}[c]{@{}c@{}} 
wikitext \\ ~\cite{wikitext}
\end{tabular} 
}
\\
\cline{2-3}
& GPT2 Large (gpt2-l)~\cite{achiam2023gpt} & 762M & \\
\cline{2-3}
& GPT2 X-Large (gpt2-xl)~\cite{achiam2023gpt} & 1.5B & \\
\cline{2-3}
& Llama 2-7B~\cite{touvron2023llama} & 7B & \\
\cline{2-3}
& Bert~\cite{kenton2019bert} & 110M & \\
\cline{2-3}
& Mixtral 8x7B~\cite{jiang2024mixtral} & 46.7B & \\
\hline
\end{tabular}
\label{tab:modeltable}
\vspace{-3mm}
\end{table}

%% file: Sections/05_CaseStudies.tex
\section{Case Studies}
\label{sec:case_studies}
% 
% \input{Tables/tab_microbench_new}
% \insertWideFigure{server_torch_energy_breakdown_Transposed}{End-to-End inference GPU energy consumption of models running on the Data Center (CPU + GPU) configuration.}
%
We conduct a thorough performance analysis of models in \autoref{tab:modeltable} on a data center and a workstation-class platform, as listed in ~\autoref{tab:experiment_settings}.
%
%The evaluated hardware platforms include a workstation and data center-class machines.
%
We employ PyTorch for our the main performance characterization and use ONNX Runtime, TensorRT, and TorchInductor for deep-dive studies (e.g., the impact of deployment flow choice and operator fusion). 

We first focuse on the GEMM and non-GEMM performance horizon (\autoref{subsec:task_dom}).
Also, we provide deeper insights into the non-GEMM performance horizon by investigating the impact of deployment flow and operator fusion (\autoref{subsec:deployment_flow_effect}), and the impact of quantization (\autoref{subsec:quantization}).

%Using the aforementioned set of deployment flows and hardware platforms on the \bench workload, we characterize the GEMM/non-GEMM performance for each model. 
%
%Using the results, we investigate the impact of the hardware platform and deployment flow.
%
%In addition, we also investigate the impact of widely adopted optimizations (operator fusion and quantization) and show that the non-GEMM performance is an issue even after applying common optimizations.

\input{Tables/tab_experiment_settings}

\subsection{Non-GEMM Performance Characterization Results}
\label{subsec:task_dom}
\insertWideFigure{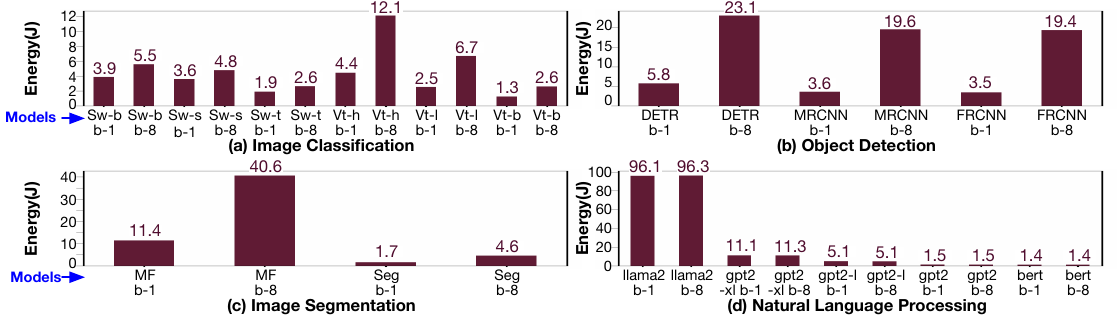}{End-to-End inference GPU energy consumption of models running on the Data Center (CPU + GPU) configuration.}
% 
%We discuss how non-GEMM operator's contribution to the end-to-end inference latency changes depending on task domains and discuss key non-GEMM operators in each domain.
We conduct a performance characterization study using PyTorch and present the results in \autoref{fig:server_op_breakdown_camera_ready}.
Aligned with what we observed in ~\autoref{fig:Intro_Motivational}, the relative contribution of non-GEMM operators to the end-to-end latency significantly increases after GEMM acceleration using a GPU, from 17.2\% to 42.3\%, on average. 
However, we observe each model show different trends, mainly affected by the non-GEMM operator types and the number of GEMM and non-GEMM operators in the original model. 
We summarize the most time-consuming non-GEMM operators in~\autoref{tab:expensive_ops_new} from the data center class platform (Platform A), which shows the diversity of the dominating non-GEMM operators in each model.
We highlight some notable models in each task category and delve into the details of their non-GEMM performance.

\betterparagraph{IC Task: Swin Transformer} For every Swin Transformer variant(Sw-t, -s, and -b), the memory operator group is the most latency-dominant non-GEMM operator group, which accounts for 32\% of the total latency, on average, on data center platform with GPU acceleration.
Those memory operators originate from the Swin Transformer's unique window shape (cross-shaped)~\cite{liu2021swin}, which is not well-aligned with memory layout of tensor data organized in dimension orders.

\betterparagraph{OD Task: DETR}
After GPU acceleration, DETR has shown significant non-GEMM operator presence, which accounts for 65.8\% of the total latency, on average.
The major source of the non-GEMM latency is in the normalization operators, whose percentages are 35\% and 32\% on Platforms A and B, respectively. 
We observe the normalization functions are based on a custom implementation, which are identified as independent kernel.
Kernel launch overheads accumulated for independent runs for the custom normalization function significantly contribute to the non-GEMM latency.
However, we also observe that an advanced deployment software (TensorRT) can fuse those operators and significantly improve the non-GEMM performance.
We discuss the details later in~\autoref{subsec:deployment_flow_effect}.

\betterparagraph{IS Task: Maskformer} MaskFormer utilizes Swin Transformer as its backbone, which introduces many memory operators as we discussed for Swin Transformer.
As a result, memory operator becomes the most dominant non-GEMM operator, which accounts for 40.8\% of the total latency, on average, as we can observe in~\autoref{fig:server_op_breakdown_camera_ready} (f).

%normalization operators are the dominant (15\% of the execution time), while Memory operators dominate the non-GEMM operators of Maskformer (24\%) as shown in~\autoref{fig:server_op_breakdown}.

\betterparagraph{NLP Task: GPT2} As we observe in~\autoref{fig:server_op_breakdown_camera_ready} (h) for both platforms, the latency of non-GEMM operators in GPT2 variants is considerable, which account for 45.0\%, on average. 
The most dominant non-GEMM operator is an activation function, GELU, which accounts for 26.4\% of the total latency.
%Normalization operators are the slowest non-GEMM operators in Llama2, running for 15\% of the total execution time as seen in~\autoref{fig:llm_onnx_vs_eager}. 

\betterparagraph{Summary} We observe GPU acceleration significantly increases the percentage of non-GEMM operators in the end-to-end latency, which amplifies the importance of non-GEMM operators in the performance optimization process. Also, we observe the most dominant non-GEMM operators are diverse depending on the model. The results indicate that an optimization technique tailored for a single operator cannot fully address the non-GEMM performance challenge, which motivates a holistic optimization approach for wide non-GEMM operators or a balanced specialization for a set of non-GEMM operators in a target workload.

\input{Tables/tab_expensive_ops_new}

\insertWideFigure {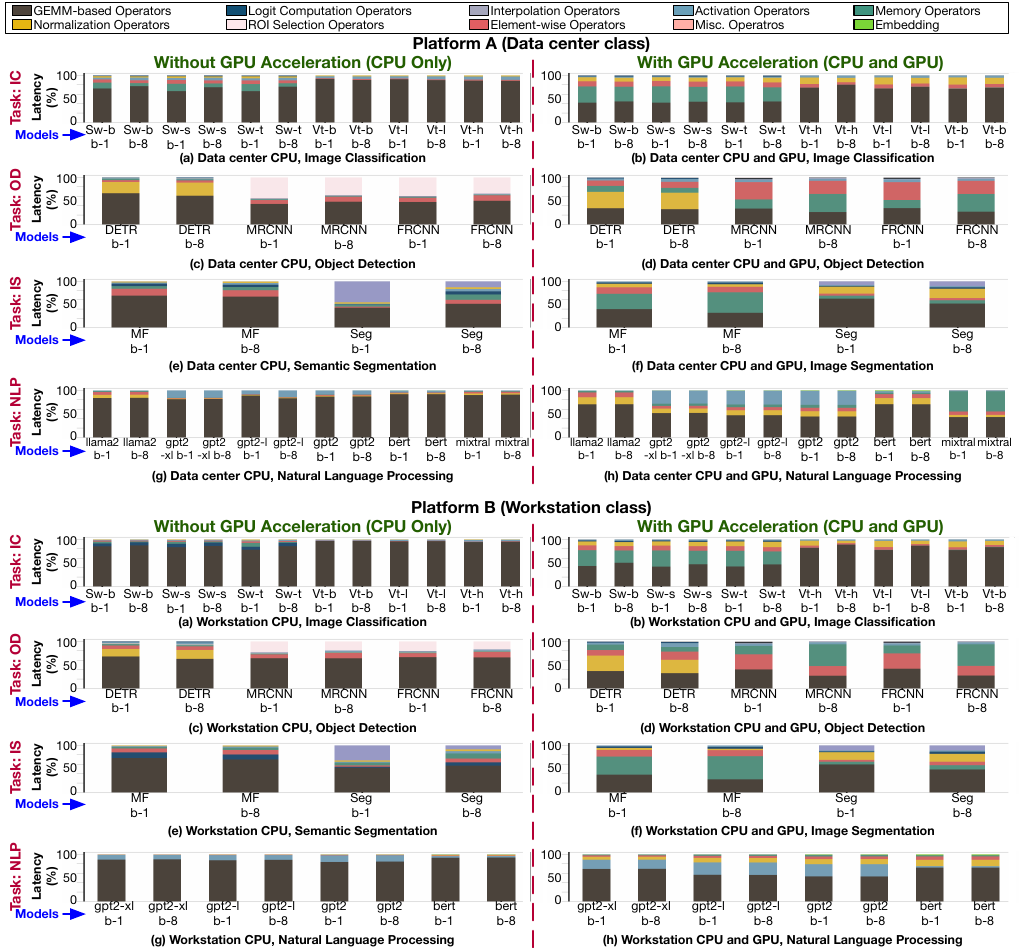}{Latency break-downs of \bench models into the operator granularity. We show CPU-only (left column) and CPU+GPU (right column) results on two evaluated platforms listed in~\autoref{tab:experiment_settings}.\vspace{-7mm}}

\subsection{The Impact of Deployment Flow on Non-GEMM Performance}
\label{subsec:deployment_flow_effect}
Deployment flows such as ONNX Runtime~\cite{onnxruntime} and TensorRT~\cite{tensorrt} are widely used for serving model inferences.
Such flows apply various optimizations to each model, which includes the computational graph optimizations (e.g., operator fusion) and backend assignment (e.g., utilizing Tensor Core in Nvidia GPUs).
To understand the impact of deployment frameworks on the non-GEMM operator performance, we conduct two case studies: (1) comparing PyTorch and ONNX Runtime (ORT) results on LLMs (focus: general optimizations w/o operator fusion) and (2) comparing PyTorch, TorchInductor, and TensorRT results (focus: operator fusion).
% how deployment frameworks affect inference performance distribution across operators, we evaluate our models with ONNX Runtime (ORT).
%

\betterparagraph{[Case Study 1] Non-GEMM Performance on LLMs across PyTorch and ORT}
%
%ORT is a cross platform ML framework that accelerates ML model inference~\cite{onnxruntime}, which leverages hardware accelerators (e.g. GPUs) and operator graph level optimizations to improve the inference performance.
%
We profile the GEMM and non-GEMM performance of two LLMs (GPT2 and Llama2) on the platform A, using the CUDA execution provider in ORT.
As the results presented in~\autoref{fig:llm_onnx_vs_eager_camera-ready} show, we observe the presence of non-GEMM operators significantly increase, from  52.6\% to a 80.75\%, on average.
We observe the percentage of memory operators significantly increases if we switch from PyTorch to ORT, from  3.2\% to 66.8\%, although the absolute end-to-end latency decreases.
Such a result originates from ORT's  significant performance boost of other operators (Lllma2) and ORT's limited efficiency on memory operators (GPT2-XL).
Many memory operators in the evaluated LLMs are not supported by the CUDA execution provider in ORT, which leads to inefficient execution on CPUs involving costly data transfer between a CPU and a GPU. 
Combined with the high frequency of such operators, the relative contribution of memory operators to the end-to-end inference increases significantly, as shown in~\autoref{fig:llm_onnx_vs_eager_camera-ready} (b).
The results imply two insights: (1) Model deployment flow can significantly aggravate the non-GEMM performance challenge and (2) the dominant non-GEMM operators differ depending on the operator support in a deployment flow.

\begin{comment}
    
\betterparagraph{Case Study 2: Non-GEMM performance with ORT across mobile and data center platforms}

We also compare the performance horizon across mobile and data center class platforms when using ORT.
%

In detection models, we notice a different trend than the remaining task domains. 
%  
The non-GEMM operator with the highest contribuion to the inference latency varies across platforms. 
% 
We notice arithmetic operators dominate 52\% of the execution when running on the mobile GPU platform, while memory operators consume 6\% of the execution time, becoming the main bottleneck on the datacenter platform. 
% 
The datacenter GPU speeds up arithmetic (the operator is NonZero, I included it under elmt-wise arithmetic) operators by 1.6$\times$. Running on the mobile GPU on the other hand, reduces the latency of these operators by a factor of 6.8$\times$, making them the major inference bottleneck on the mobile GPU. 
% 
This indicates that HW platform will affect the performance of non-GEMM operators.   
% 

\end{comment}

\insertFigure{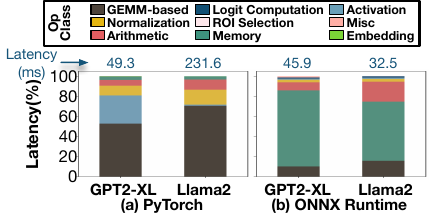}{The impact of deployment software toolchain into the latency breakdown on language models. (a) PyTorch and (2) ONNX Runtime on a data center class GPU (A100).}
% The numbers at the top of each bar represent the raw latencies in milliseconds (ms).}

%\insertFigure{onnx_hw_comparison}{\TODO{Add Workstation to the plot}The latency breakdown of the inference using ORT on two platform configurations: mobile and Data Center.}

\insertFigure{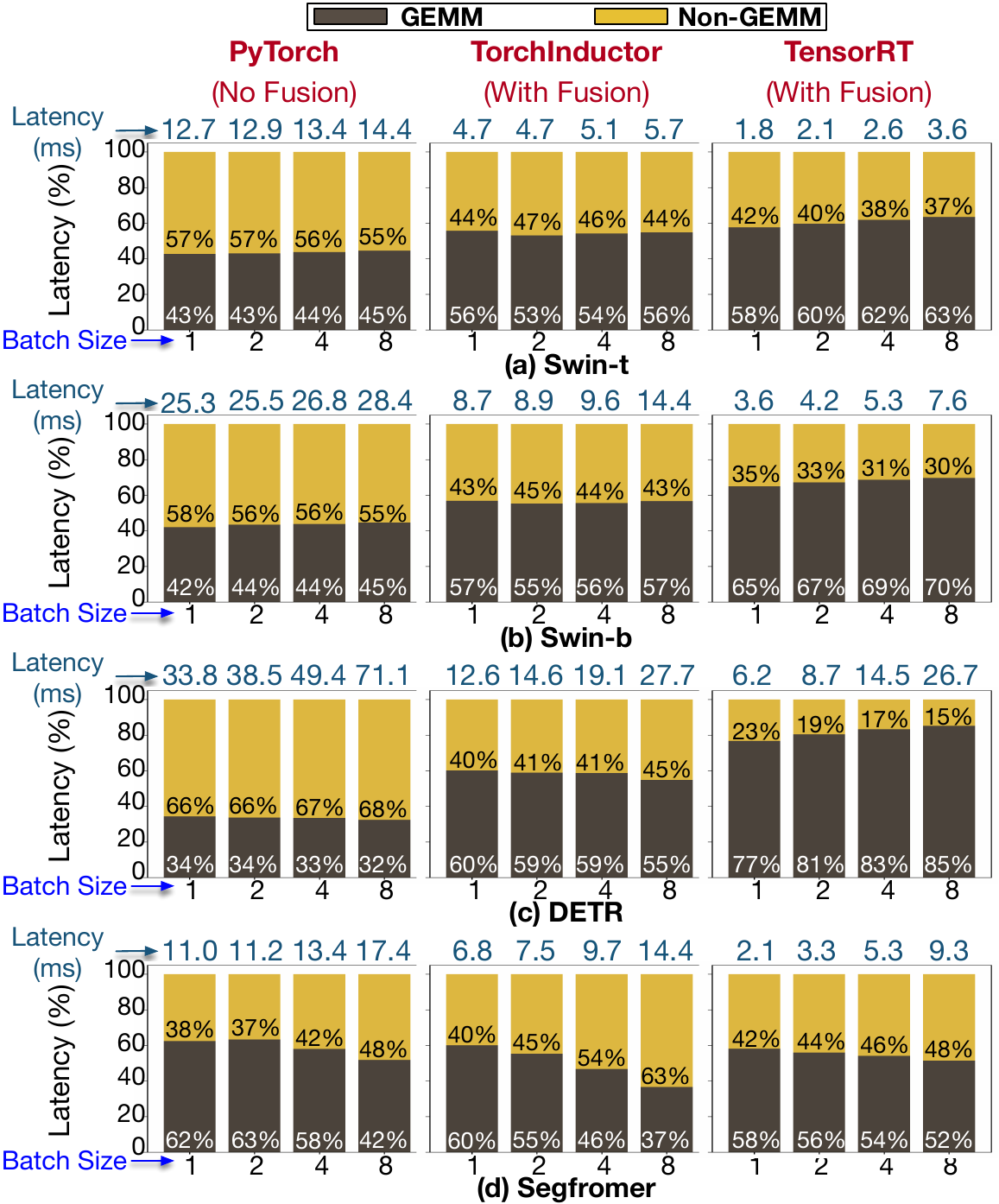} {The latency of non-GEMM operator in (a) Swin-t, (b) Swin-b, (c) DETR, and (d) Segformer is still considerable despite applying operator fusion in TorchInductor (middle) and TensorRT (right).}
% The numbers at the top of each bar represent the latencies in milliseconds.}

% \insertFigure{server_trt_lm}{Non-GEMM operators contribution to end-to-end LLM inference latency is still considerable despite applying operator fusion.  For (a) GPT2-XL and (b) Llama2, non-GEMM latency increases with the sequence length when running in TensorRT.}

\betterparagraph{[Case Study 2] Non-GEMM Performance with Operator Fusion}
%\label{subsubsec:fusion}
%
% \TODO{Swin-b data in Fig.9 shows that even with optimizations applied to non-GEMM operators, non-gemm is still considerable}
%
% \TODO{Look at the swin-tiny TRT vs Pytorch gemm non gemm and speed up. We want to verify the claim that models like segformer and swin-tiny were dominated by non-compute operations}
%
% \TODO{Main point we are showing is the effect of the fusion, We can notice that normalization is fusion friendly, while other operators in the Segformer}
Operator fusion is one of the key optimization technique for accelerating inference workloads~\cite{dao2022flashattention, korch, symons2023streamfuse, niu2021dnnfusion, tensorrt, kao2023flat}. 
Operator fusion combines multiple operators in a single kernel to reduce the number of costly kernel launches and minimize the number of redundant offchip memory accesses~\cite{niu2021dnnfusion}.
%input reads and intermediate tensor stores~\cite{niu2021dnnfusion}. 
% 
TensorRT~\cite{tensorrt} is a widely adopted inference framework  released by Nvidia that applies the operator fusion technique targeting GPUs.
Operator fusion in TensorRT detects specific patterns (e.g., three consecutive element-wise operators~\cite{tensorrt}) in the operator graph and fuses nodes captured in the patterns to enhance inference performance by reducing redundant memory accesses around non-GEMM operators. 
% 
%Common fusion patterns include combining multiple element-wise non-GEMM operators or integrating non-GEMM operators with GEMM operators to alleviate latency bottlenecks.
% 
%Typically, operator fusion is applied to reduce non-GEMM operator bottlenecks when accelerating the inference on GPU. 
%  

% Although operator fusion is a technique typically applied to reduce bottlenecks of non-GEMM operators that get amplified by GEMM acceleration on GPU, we notice in~\autoref{fig:server_trt_cv} that fusion does not fully address non-GEMM operators bottlenecks for all models.
% 
% 

%\betterparagraph{Impact of Operator Fusion} 

To understand the impact of operator fusion on the non-GEMM performance, we conduct a case study on four models listed in~\autoref{fig:server_trt_cv_inductor_camera_ready}, comparing TensorRT (with fusion) and PyTorch (without fusion).
%and compare the GEMM/Non-GEMM latency breakdowns of the models using TensorRT (with fusion) and PyTorch (without fusion).
%
We present the results in~\autoref{fig:server_trt_cv_inductor_camera_ready}, which shows the inference latency breakdown between GEMM and non-GEMM operators on Platform A (data center class).
The results indicate that fusion mitigates the non-GEMM bottleneck, but it does not completely address the challenge. 
For example on Swin-b, after applying operator fusion by switching to TensorRT from PyTorch, the contribution of non-GEMM operators to the total latency changes from 56.4\% to 32.2\%, on average.
The reduction in the percentage is based on the non-GEMM performance improvement via operator fusion, which reduces 88.6\% of latency, as summarized in~\autoref{tab:fusion_rate}.
However, the non-GEMM operators still account for 32.2\% of total latency.
This shows that operator fusion cannot eliminate the non-GEMM performance challenge and motivates further studies toward non-GEMM performance optimization.

%before applying fusion, non-GEMM operators dominate Swin-b inference latency, accounting for 58\% of the total.

%After Applying operator fusion with TensorRT speeds up the Swin-b non-GEMM latency by 8.84$\times$ as indicated in~\autoref{tab:fusion_rate}. 
%  
%However, non-GEMM operators still contribute a considerable 39\% of the total inference latency. 
% 
%\textit{This indicates that fusion does not fully mitigate the non-GEMM bottleneck}.
% 

%\betterparagraph{Factors Affecting the Effectiveness of Fusion}
Although most results indicate a considerable impact of non-GEMM even after operator fusion, we observe that TensorRT operator fusion on the DETR model is exceptionally effective.
Therefore, we conduct a deep-dive study, investigating the percentage of fused non-GEMM operators (i.e., fusion rate) and performance improvements after fusion, as listed in~\autoref{tab:fusion_rate}.
We observe the strong non-GEMM performance improvements for DETR originates from high fusion rate of 30\%, which led to 13.5$\times$ non-GEMM speedup.
This leads to the large percentage reduction of non-GEMM in the total latency, from 66.5\% to 18.5\%, on average.

However, the fusion rate is not the only factor that determines the non-GEMM speedup.
For example, DETR and Segformer have similar fusion rates (30\% and 27\%, respectively), but the amount of non-GEMM performance improvements are significantly different: 13.5$\times$ and 2.39$\times$, respectively. 
We analyze the execution trace and identify the fusion pattern around batch normalizations as the main source of the difference.
Most batch normalization operators (100\% of total) in DETR were fused with GEMM-operators (CONV+BN+ReLu pattern) while those in Segformer were fused with other non-GEMM operators (97.8\% of total). 
The results indicate that the effectiveness of operator fusion relies on not only the overall fusion rate but also the fusion patterns.
%DETR: 84 ops 
%SegFormer: 31 ops 
%
%Note that the effective fusion pattern in DETR originates from the convolutional neural network used as backbone: as Transformer-based models are trending to be state-of-the-art, we cannot expect DETR-like improvements in general.
%
Our observation confirms that the operator fusion cannot fully address the non-GEMM performance challenge, even if it can be very effective on some patterns.
\input{Tables/tab_fusion_rate}

\subsection{The Impact of Quantization Non-GEMM Performance}
\label{subsec:quantization}

Quantization refers to the model optimization technique, which reduces the bit precision of model weights and/or activations to enhance the computational performance and efficiency of DNN inference.
Quantization is a widely-adopted technique~\cite{jacob2018quantization, nagel2021whitequant} including heavy models like LLMs~\cite{frantar2022gptq, lin2024awq, xiao2023smoothquant, llm.int8}.
%
%For example, LLM.int8()~\cite{llm.int8} introduced an 8-bit - 16-bit mixed precision quantization scheme for LLMs.
%
LLM.int8() is a state-of-the-art quantization method, which quantizes more than 99\% of the linear layers in OPT LLM to an 8-bit precision.
Therefore, we adopt LLM.int8() and characterize GEMM and non-GEMM performance of Llama3 on Platform A to understand the impact of quantization on the non-GEMM operator performance problem.

As the results in~\autoref{fig:server_quantization_camera_ready} show, we observe that non-GEMM operators dominate the latency after quantization, changing from 29.3\% to 76.7\%, on average.
Such a significant shift in the latency distribution is mainly based on the GEMM performance improvements from 8-bit arithmetic, which reduced the latency by 38.2\%, on average.
However, based on our analysis on the execution traces, non-GEMM performance aggravates because the 8-bit data need to be dequantized and re-quantized for non-GEMM operators, which requires 16-bit floating point arithmetic.
This introduced 6510 additional non-GEMM operators into the computation graph, which led to a significant increase of the non-GEMM latency by 5.6$\times$ after quantization.
%q
Combined together, the overall percentage of non-GEMM operators in the total latency dominate after quantization, which makes non-GEMM operators as the major optimization target.

In the case study on Llama3 8B, we observe the element-wise arithmetic operators originate from dequantization/requantization (DQRQ) process dominate in the inference latency, which adds 20\% extra non-GEMM operators to the original computational graph. 
Also, we observe longer sequence length leads to higher percentage in the element-wise arithmetic operators.
For example, as we increase the sequence length from 512 to 8192, the latency percentage of element-wise arithmetic operators increase from 31.8\% to 63.8\%.
As current trends in the LLMs are toward longer sequence lengths~\cite{dubey2024llama3, achiam2023gpt}, the non-GEMM performance issues in longer sequences originating from DQRQ costs will aggravate, which motivates efforts in non-GEMM performance optimization.

\insertFigure{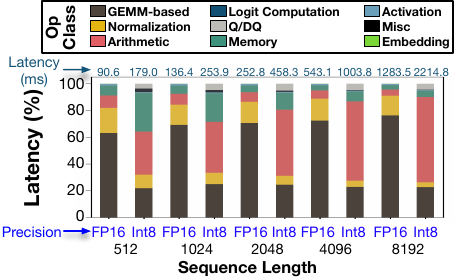}{GEMM and Non-GEMM latency breakdown of the inference latency on Platform A, running an 8-bit quantized Llama3 8B~\cite{llm.int8}.}

% The numbers at the top of each bar represent the raw latencies in milliseconds (ms).}

% \input{Tables/tab_fusion_rate}

% \insertWideFigure{server_torch_energy_breakdown_Transposed}{End-to-End inference GPU energy consumption of models running on the Data Center (CPU + GPU) configuration.}
% 
% \input{Tables/tab_energy}
\subsection{Key Observations and Insights}
\label{subsec:observations}

We summarize our main observations and insights:

\begin{itemize}
{\item After GEMM acceleration, non-GEMM becomes a major optimization target regardless of the hardware platforms.}
{\item Specialized optimization for one non-GEMM is not effective due to the diversity in dominant non-GEMM.}
{\item Operator fusion cannnot fully address non-GEMM performance challenge: Although it can significantly improve non-GEMM performance, but its effectiveness heavily depends on the model.}
{\item Operator support in deployment flows significantly affects the non-GEMM performance.
} 
%}
{\item Quantization significantly aggravates the non-GEMM performance challenge due to the imbalanced speedup across GEMM and non-GEMM and quantization/dequantization costs around non-GEMM operators
}
\end{itemize}

%% file: Tables/tab_experiment_settings.tex
% \begin{center}
% \begin{table}
% \centering
% \scriptsize
% \caption{Hardware platform configurations used for \bench case studies.}
% \begin{tabular} {| l | l | l |}
% \hline
% \multicolumn{1}{|c|}{\textbf{Class}} & \multicolumn{1}{|c|}{\textbf{CPU}} & \multicolumn{1}{|c|}{\textbf{GPU}} \\
% \hline 
% Data Center & AMD EPYC 7763 & NVIDIA A100 \\
% \hline
% Workstation & Intel i9-13900K & NVIDIA RTX 4090 \\
% \hline
% Mobile & Intel i7-13700HU & NVIDIA RTX 4060m \\
% \hline
% \end{tabular}
% \label{tab:experiment_settings}

% \end{table}
% \end{center}

\begin{table}[]
\centering
\scriptsize
\caption{Hardware platform configurations used for case studies.}
\begin{tabular}{|l|l|l|l|l|l|}
\hline
\multicolumn{1}{|c|}{\multirow{2}{*}{\textbf{ID}}} 
& \multicolumn{1}{|c|}{\multirow{2}{*}{\textbf{Class}}} 
& \multicolumn{1}{c|}{\multirow{1}{*} {\textbf{CPU}}}             
& \multicolumn{3}{c|}{\multirow{1}{*}{\textbf{GPU}}}
\\ 
\cline{3-6}
&
& \multicolumn{1}{c|}{\textbf{Device}}             
& \multicolumn{1}{c|}{\textbf{Device}}             
& \multicolumn{1}{c|}{\textbf{Mem.}}
& \multicolumn{1}{c|}{\textbf{TOPS}} \\ 
\hline
A
& Data Center                                   
& \multicolumn{1}{l|}{AMD EPYC 7763}                           
& \multicolumn{1}{l|}{Nvidia A100}      
& \multicolumn{1}{r|} {80 GB}
& \multicolumn{1}{r|} {624} \\ 
\hline
B
& Workstation                                  
& \multicolumn{1}{l|}{Intel i9-13900K}                         
& \multicolumn{1}{l|}{Nvidia RTX 4090}  
& \multicolumn{1}{r|} {24 GB}
& \multicolumn{1}{r|} {660} \\ 
\hline
\end{tabular}
\vspace{-5mm}
\label{tab:experiment_settings}
\end{table}

%% file: Tables/tab_expensive_ops_new.tex
\begin{table}
\caption{Most time-consuming non-GEMM operator groups for selected Models (Platform A, with GPU acceleration, average across batch sizes).}

\scriptsize
\centering

\begin{tabular} {   | c | c | c | c |  } 
\hline

\multirow{2}{*}{
\begin{tabular}[c]{@{}c@{}} 
\textbf{Task} \\ \textbf{Domain}
\end{tabular}} 
&
\multirow{2}{*}{
\textbf{Model}}
& 

\multirow{2}{*}{
\begin{tabular}[c]{@{}c@{}} 
\textbf{Operator}  \\ \textbf{Group}  
\end{tabular}} 
&
\multirow{2}{*}{
\begin{tabular}[c]{@{}c@{}} 
\textbf{Latency} \\ \textbf{Percentage (\%)}  
\end{tabular}} 
\\

&&& \\

% \textbf{ Application } & \textbf{ Models } & \textbf{Batch} & \textbf{ Operator Group } & \textbf{\% of Exec.} 
%
%  &   &  \textbf{Size}& \textbf{Group }   & \textbf{Time }  \\
\hline
\multirow{6}{*}{
{
\begin{tabular}[c]{@{}c@{}} 
Image  \\ Classification
\end{tabular}}
}
&\multirow{1}{*}{
{
\begin{tabular}[c]{@{}c@{}} 
 Vt-b    
\end{tabular}}
}  
 &  Norm  & 14.0 \\ 
\cline{2-4}
& \multirow{1}{*}{
{
\begin{tabular}[c]{@{}c@{}} 
Vt-l 
\end{tabular}}}  & Norm & 13.3 \\ 
\cline{2-4}
& \multirow{1}{*}{
{
\begin{tabular}[c]{@{}c@{}} 
Vt-h 
\end{tabular}}}  & Norm & 11.2 \\ 
\cline{2-4}
& \multirow{1}{*}{
{
\begin{tabular}[c]{@{}c@{}} 
 Sw-t  
\end{tabular}}} & Memory & 31.8 \\ 
\cline{2-4}

&\multirow{1}{*}{
{
\begin{tabular}[c]{@{}c@{}} 
Sw-s  
\end{tabular}}} & Memory & 33.1 \\ 
\cline{2-4}

&\multirow{1}{*}{{
\begin{tabular}[c]{@{}c@{}} 
Sw-b   
\end{tabular}}}  & Memory & 32.8 \\ 
\hline
\multirow{3}{*}{Object Detection} 
&  \multirow{1}{*}{FRCNN}
& Elmt-wise Arith. & 34.4 \\ 
\cline{2-4}
&  \multirow{1}{*}{MRCNN}  & Elmt-wise Arith. & 33.6\\
\cline{2-4}
& DETR & Norm & 34.8 \\ 
\hline
\multirow{2}{*}{Image Segmenation} 
&  \multirow{1}{*}{MF}
& Memory & 40.8 \\ 
\cline{2-4}
&  \multirow{1}{*}{Seg}  & Normalization & 17.4\\
\hline
\multirow{5}{*}{
{
\begin{tabular}[c]{@{}c@{}} 
 NLP \\  
\end{tabular}}} 
& \multirow{1}{*}{
{
\begin{tabular}[c]{@{}c@{}} 
gpt2  
\end{tabular}}} & Act & 30.2 \\
\cline{2-4}
&\multirow{1}{*}{{
\begin{tabular}[c]{@{}c@{}} 
GPT2-L
\end{tabular}}}  & Act & 29.9 \\
\cline{2-4}
&\multirow{1}{*}{{
\begin{tabular}[c]{@{}c@{}} 
GPT2-XL
\end{tabular}}} & Act & 28.1 \\
\cline{2-4}
& Llama2 &   Norm & 14.9 \\
\cline{2-4}
&\multirow{1}{*} {bert} &   Norm & 13.1 \\
\cline{2-4}
&\multirow{1}{*} {Mixtral} &   Memory & 43.1  \\
\hline 
\end{tabular}
\label{tab:expensive_ops_new}
\vspace{-0mm}
\end{table}

%% file: Tables/tab_fusion_rate.tex
\begin{table}[t]
\caption {The non-GEMM latency before and after applying fusion with TensorRT. The values between brackets represent the percentage with resect to the total inference latency.}

\centering

\begin{tabular}{|l|l|l|l|}
\hline
\multicolumn{1}{|c|}{\multirow{2}{*}{\textbf{Model}}} 
& \multicolumn{1}{c|}{\multirow{1}{*}{\textbf{Non-GEMM}}} 
& \multicolumn{2}{c|}{\multirow{1}{*}{\textbf{Non-GEMM Latency}}} \\ 
\cline{3-4}
& \multirow{1}{*}{\textbf{Fusion Rate}}
& \multirow{1}{*}{\textbf{Before Fusion}} 
& \multirow{1}{*}{\textbf{After Fusion}} \\
\hline

\multicolumn{1}{|c|}{Swin-t}  
& \multicolumn{1}{r|}{8.8\%}
& \multicolumn{1}{r|}{7.53 ms (\textit{56.4\%})}
& \multicolumn{1}{r|}{0.97 ms (\textit{39.0\%})} \\ 
\hline
\multicolumn{1}{|c|}{Swin-b}  
& \multicolumn{1}{r|}{7.0\%}
& \multicolumn{1}{r|}{14.59 ms (\textit{56.4\%})}
& \multicolumn{1}{r|}{1.65 ms (\textit{32.3\%})} \\ 
\hline
\multicolumn{1}{|c|}{DETR}  
& \multicolumn{1}{r|}{30.0\%}
& \multicolumn{1}{r|}{32.17 ms (\textit{66.4\%})}
& \multicolumn{1}{r|}{2.38 ms (\textit{18.5\%})} \\ 
\hline
\multicolumn{1}{|c|}{Segformer}  
& \multicolumn{1}{r|}{27.0\%}
& \multicolumn{1}{r|}{5.57 ms (\textit{41.0\%})}
& \multicolumn{1}{r|}{2.33 ms (\textit{41.0\%})} \\ 
\hline
\end{tabular}
\label{tab:fusion_rate}
\vspace{-2mm}
\end{table}

%% file: Sections/06_RelatedWorks.tex
\section{Related Works}
\label{sec:related_works}

\betterparagraph{ML Inference Benchmarks}Many end-to-end inference benchmarks~\cite{hao2023torchbench, reddi2020mlperf} do not capture operator level performance breakdowns.
MLPerf Inference~\cite{reddi2020mlperf}, an industry standard inference benchmark, offers a flexible and standardized framework to evaluate the performance of inference systems.
MLPerf Inference framework defines performance metrics and workloads, and supports measuring the performance of realistic inference scenarios across a wide range of software and hardware systems.
Nevertheless, MLPerf does not offer any operator-level fine-grained latency breakdowns, which makes it unsuitable for understanding the implications of non-GEMM operators on the inference performance. 
Our work provides fine-grained operator level latency breakdowns to understand the impact of non-GEMM operators on the end-to-end inference performance.
\betterparagraph{Non-GEMM Characterization} Previous works~\cite{longtailbench, pati2023tale,choi2022accelerating, stevens2021softermax, wang2021spatten, ghodrati2024tandem} investigate non-GEMM operators, however their characterization and optimization focus on a limited set of operators and applications.
Longtail bench~\cite{longtailbench} proposes a microbenchmark specific to a limited set of non-GEMM operators from selected computer vision models. 
It profiles operators without a compute library implementation in a standalone setting, using randomly generated data. 
% 
%Despite being useful to identify bottlenecks in custom kernels,
Because Longtail bench is a microbenchmark suite, it cannot be used to capture realistic interplay between GEMM and non-GEMM operators in real models, which provides insights for inter-operator optimizations. 
In addition, Longtail bench does not provide general insights on non-GEMM performance because it focuses on a specific computer vision application.
Our work extends on these efforts by studying the non-GEMM performance of 17 popular models in realistic end-to-end inference scenarios covering various task domains.
Tandem Processor~\cite{ghodrati2024tandem} highlights the importance of non-GEMM operator-oriented optimization in ML inference, and proposes a co-processor architecture to mitigate the non-GEMM overhead.  
Tandem Processor characterized the non-GEMM performance in 7 models and identify non-GEMM operators as the emerging bottleneck after accelerating GEMMs. 
Although Tandem is a pioneering work in non-GEMM optimization, our work provides additional and broader insights beyond it. 
Our work evaluates 17 widely adopted models across task domains and offers detailed case studies analyzing the impact of widely adopted optimization techniques, operator fusion and quantization, on the non-GEMM performance.

%% file: Sections/07_Conclusion.tex
\section{Conclusion}
\label{sec:conclusion}
Accelerating GEMM operators in ML inference have changed the major bottleneck from GEMM to non-GEMM operators. %to have more presence of non-GEMM operators in the performance horizon.
To understand the latest GEMM/non-GEMM performance landscape with GEMM acceleration, we conducted a thorough performance analysis of non-GEMM operators in the latest models in various task domains and platforms.
The results confirm the increasing importance of non-GEMM performance and show that common model optimizations (e.g., quantization) can significantly aggravate the non-GEMM performance challenge.
The dominant non-GEMM operators are diverse across models, which indicates that a specialized optimization targeting a specific operator cannot solve the non-GEMM performance challenge.
We also show that non-GEMM-oriented optimization such as operator fusion cannot fully address the non-GEMM performance challenge.

Our performance anlysis results imply that now we need to consider non-GEMM operators as a major optimization target and develop new hardware and software techniques to optimize non-GEMM performance.
The non-GEMM profiling software we used in this study, \bench, will be released as open-source software, which will contribute to the follow-up research for non-GEMM optimization.

%To guide future optimization, we need to understand the performance of non-GEMM operators in recent popular models. 
%
%In this work, we analyze the performance non-GEMM operators in recent popular machine learning models.  
%
%We show that the dominance of non-GEMM operators is occurring in all platforms with GPUs, and the deployment software's impact is large.
%
%We also highlight that common optimizations like operator fusion does not fully address the non-GEMM bottlenecks.
% % 
%We construct \bench to study of non-GEMM operators in accelerated popular models reflecting realistic use scenarios and evaluate it on different hardware platforms.
% 
%Our findings strongly motivate ML system researchers to consider non-GEMM operators as a major consideration.
%
%We believe \bench sheds light on new potential optimization directions to improve the performance of ML systems.

%We evaluate \bench on different hardware platforms to offer the community insights on the performance of accelerated ML workloads and new potential optimization opportunities. 
% 

%% file: Sections/09_Appendix_A.tex
\clearpage
\appendices
\section*{Artifact Appendix}
%%%%%%%%%%%%%%%%%%%%%%%%%%%%%%%%%%%%%%%%%%%%%%%%%%%%%%%%%%%%%%%%%%%%%
\subsection{Abstract}
This appendix describes the workflow to run \bench and reproduce the results reported in the paper.

\subsection{Artifact check-list (meta-information)}

{\small
\begin{itemize}
  \item {\textbf{Algorithm}: Profiling functions are deployment flow specific. We use the PyTorch Profiler~\cite{pytorch_profiler} for PyTorch, EP Profiling for ONNX RUNTIME~\cite{onnxruntime}, and TensorRT Open Source Software (OSS) for TensorRT~\cite{trt_oss}. }
  \item {\textbf {Program}: Python 3.10, CUDA 12.6.}
  \item {\textbf{Model:} Please refer to~\autoref{tab:modeltable}.}
  \item {\textbf {Data set:} Please refer to~\autoref{tab:modeltable}. }
  \item {\textbf {Run-time environment:} Tested Environments: Ubuntu 22.04, Linux Mint 21.1, and MacOS 14.2.1.}
  \item {\textbf {Hardware:} AMD EPYC 7763 CPU, 1TB DDR4 RAM, 1 x NVIDIA A100 80GB (PCIe), Intel i9-13900K, 64 GB DDR5 RAM, and Nvidia RTX 4090 24GB (PCIe).  }
  \item {\textbf {Execution:} Automated Scripts. Please refer to the README file in the Github repository. }
  \item {\textbf {Metrics:} Latency.}
  \item {\textbf {Output:} Plots in PNG format, and the corresponding data in csv format. The automated scripts plot the operator level end-to-end inference latency breakdown of all \bench profiled models.}
  \item {\textbf {Experiments:} Please refer to \autoref{sec:case_studies} for more details.  }
  \item {\textbf {How much disk space required (approximately)?:} Approximately 100 GB to store the models, the datasets, and the collected profiling traces.}
  \item {\textbf {How much time is needed to prepare workflow (approximately)?:} Approximately, setting up the workflow requires around 30 minutes. }
  \item {\textbf {How much time is needed to complete experiments (approximately)?:} Approximately 10 hours.}
  \item {\textbf {Publicly available?:} https://doi.org/10.5281/zenodo.15043135}
  \item {\textbf {Code licenses (if publicly available)?:}  MIT License. }
  \item {\textbf {Archived (provide DOI)?:} 10.5281/zenodo.15043135 }
\end{itemize}
}

%%%%%%%%%%%%%%%%%%%%%%%%%%%%%%%%%%%%%%%%%%%%%%%%%%%%%%%%%%%%%%%%%%%%%
\subsection{Description}

\subsubsection{How to access}
The source code is available on Zenodo at   https://doi.org/10.5281/zenodo.15043135, or on Github at https://github.com/UCI-ISA-Lab/NonGEMM-Bench-ISPASS25.git. 

\subsubsection{Hardware dependencies}
To reproduce the paper's results, the following systems are required: 
\begin{itemize}
\item{Server with an AMD EPYC 7763 CPU, 1TB DDR4 RAM, 1x Nvidia A100 80GB GPU. (We note that the Mixture of Experts model profiling requires 4x Nvidia A100 80GB GPUs.) }
\item{Workstation with an Intel i9-13900K, 64 GB DDR5 RAM, 1x Nvidia RTX 4090 24 GB GPU.}
\end{itemize}
Nevertheless, our workflow runs on any typical laptop, workstation, or server system with a CUDA-capable GPU. 

\subsubsection{Software dependencies}
\begin{itemize}
\item{Python 3.10}
\item{CUDA 12.6}
\item{TensorRT 10.4.0.26}
\item{TensorRT Open Source Software}
\item{PyTorch}
\item{Torchvision}
\item{ONNX Runtime}
\item{Hugging Face Transformers}
\item{Hugging Face Datasets}
\item{Hugging Face Optimum}
\item{Hugging Face Accelerate}
\item{Matplotlib}
\item{COCO API}
\item{Access to Llama 2 Weights on Hugging Face}
\item{Access to Llama 3 Weights on Hugging Face}
\item{Access to Mixtral 8x7B Weights on Hugging Face}
\end{itemize}

\subsubsection{Data sets}
We use three publicly available datasets highlighted in~\autoref{tab:modeltable}: ImageNet~\cite{imagenet_cvpr09}, COCO~\cite{lin2014microsoft}, and wikitext~\cite{wikitext}

\subsubsection{Models}
We use 17 popular pretrained models from Huggingface and Torchvision. Please refer to~\autoref{tab:modeltable} for the detailed list.

%%%%%%%%%%%%%%%%%%%%%%%%%%%%%%%%%%%%%%%%%%%%%%%%%%%%%%%%%%%%%%%%%%%%%
\subsection{Installation}
\subsubsection {PyTorch and ONNX Runtime Flow Software Dependency Installation}

\begin{verbatim}
> cd torch_flow
> conda create -n ng-torch python=3.10
> pip install -r requirements.txt 
> ## After setting up the conda environment, 
> ## Install the COCO dataset dependencies.
\end{verbatim}
Please refer to the code base for more details. 

\subsubsection{TensorRT Flow Software Dependency Installation}

\begin{verbatim}
> cd trt_flow
> conda create -n ng-trt python=3.10
> pip install -r requirements.txt 
\end{verbatim}
After setting up the conda environment, please refer to the TensorRT OSS Github repository (https://github.com/NVIDIA/TensorRT/tree/release/10.4) to setup TensorRT. 

%%%%%%%%%%%%%%%%%%%%%%%%%%%%%%%%%%%%%%%%%%%%%%%%%%%%%%%%%%%%%%%%%%%%%
\subsection{Experiment workflow}
\begin{verbatim}
> cd torch_flow
> conda activate ng-torch
> ## Set the path to ImageNet and COCO datasets in run.py
> bash run_ispass_all.sh
> cd ../onnx_flow 
> bash run_ispass_all.sh
> cd ../trt_flow 
> conda activate ng-trt
> ## Set the path to your TensorRT OSS
> ## installation in setup.sh
> bash run_ispass_all.sh

\end{verbatim}

\textbf{Note:} Before running the experiments, environment variables and global constants should be properly set to configure the path to the datasets and to the TensorRT tools. Please refer to the README file in the codebase.
%%%%%%%%%%%%%%%%%%%%%%%%%%%%%%%%%%%%%%%%%%%%%%%%%%%%%%%%%%%%%%%%%%%%%
\subsection{Evaluation and expected results}

Running the \texttt{run\_ispass\_all.sh} scripts in every subdirectory will reproduce~\autoref{fig:server_op_breakdown_camera_ready}, ~\autoref{fig:llm_onnx_vs_eager_camera-ready} , ~\autoref{fig:server_trt_cv_inductor_camera_ready}, and ~\autoref{fig:server_quantization_camera_ready}. 

The scripts will generate the plots and the corresponding CSV data in the \texttt{torch\_flow/summary}, \texttt{onnx\_flow/fig6\_onnx}, and \texttt{trt\_flow/fig7\_trt}. The raw data is stored in \texttt{non-gemm-out} directory. 
The reproduced latency results are expected to be close to the results in the paper, but not an exact match because of potential differences in the hardware or software environment. 
% 

%%%%%%%%%%%%%%%%%%%%%%%%%%%%%%%%%%%%%%%%%%%%%%%%%%%%%%%%%%%%%%%%%%%%%
\subsection{Experiment customization}
The provided scripts run the entire evaluation presented in the paper. 
The users can customize their experiments by modifying the following files: 
\begin{itemize}
\item{\textbf{Modifying datasets and profiling settings:} Please modify corresponding variables in \texttt{torch\_flow/run.py}.}
\item{\textbf{Profiling new models:} Please refer to \texttt{ModelProfile} class in \texttt{torch\_flow/run.py} and add the desired model to the file.}

\end{itemize}

%%%%%%%%%%%%%%%%%%%%%%%%%%%%%%%%%%%%%%%%%%%%%%%%%%%%%%%%%%%%%%%%%%%%%
\subsection{Methodology}

Submission, reviewing and badging methodology:

\begin{itemize}
  \item \url{https://www.acm.org/publications/policies/artifact-review-and-badging-current}
  \item \url{https://cTuning.org/ae}
\end{itemize}